\documentclass[final,5p,times,twocolumn]{elsarticle}

\usepackage{amsmath}
\usepackage{amssymb}
\usepackage{graphicx}
\usepackage{bm}
\usepackage[utf8]{inputenc}
\usepackage[T1]{fontenc}
\usepackage{cleveref}
\usepackage{xcolor}
\usepackage{booktabs}

\biboptions{numbers,sort&compress,square}

\journal{Journal of Computational Physics}

\newcommand{\vE}{\bm{E}}          
\newcommand{\vB}{\bm{B}}          
\newcommand{\vA}{\bm{A}}          
\newcommand{\vJ}{\bm{J}}          
\newcommand{\vJi}{\bm{J}_i}       
\newcommand{\vJe}{\bm{J}_e}       
\newcommand{\vx}{\bm{x}}          
\newcommand{\vv}{\bm{v}}          
\newcommand{\curl}{\nabla\times}
\newcommand{\diver}{\nabla\cdot}
\newcommand{\grad}{\nabla}
\newcommand{\pe}{P_e}             
\newcommand{\nel}{n_e}            
\newcommand{\Te}{T_e}             
\newcommand{\etah}{\eta_h}        
\newcommand{\dt}{\Delta t}
\newcommand{\ntheta}[1]{#1^{n+\theta}}   
\newcommand{\nhalf}[1]{#1^{n+1/2}}       
\newcommand{\vVe}{\bm{V}_e}           
\newcommand{\vJext}{\bm{J}_\mathrm{ext}} 
\newcommand{\DV}{\Delta V}            
\newcommand{\vD}{\bm{D}}              
\newcommand{\vEstar}{\vE^{\ast}}      

\graphicspath{{Figures/}}

\begin{document}

\begin{frontmatter}

\title{A $\theta$-implicit hybrid particle-in-cell algorithm for kinetic simulations of magnetized plasmas}

\author[helion]{Prabhat Kumar\corref{author}}
\author[helion]{William J. White}
\author[helion]{Stephen E. Clark}

\cortext[author]{Corresponding author.\\
\textit{E-mail address:} prabhat.kumar@helionenergy.com}

\address[helion]{Helion Energy, Inc., Everett, WA, USA}

\begin{abstract}
Hybrid particle-in-cell (PIC) models with kinetic ions and an
inertialess electron fluid are widely used to simulate magnetized
plasmas at ion scales, but explicit time advances are constrained by the
whistler-wave CFL condition $\dt \propto \Delta x^2$, and the
sub-cycling and artificial dissipation used to mitigate it carry
accumulating costs. We present a $\theta$-implicit hybrid PIC algorithm,
implemented in the open-source GPU-capable code WarpX, in which fields
and particles are advanced as a single nonlinear system solved by a particle-suppressed 
Jacobian-free Newton--Krylov method: the particles are nonlinearly
eliminated, the electric field of the generalized Ohm's law is the sole
unknown, and mass matrices remove all particle operations from the
linear iterations. The electron pressure is advanced inside the
nonlinear solve, and the scheme satisfies a discrete energy law whose
single dissipation term vanishes at $\theta = 1/2$ and damps only
temporally under-resolved modes for $\theta > 1/2$. The advance
reproduces parallel and perpendicular wave dispersion including the ion
Bernstein modes, conserves total energy to machine roundoff where the
explicit advance heats secularly, does not develop the cold-beam
finite-grid instability, and reproduces the $m=0$ instability of a
Bennett pinch in RZ geometry. We conclude with 3D simulations of a
field-reversed configuration initialized from a Grad--Shafranov
equilibrium, comparing the explicit and implicit advances on the growth
of the tilt instability and the preservation of trapped magnetic flux.
\end{abstract}


\end{frontmatter}

\section{\label{sec:intro}Introduction}

Fusion-relevant, compact high-beta magnetized plasmas evolve on timescales that span from ion kinetic motion to
global transport. A prominent example is the field-reversed
configuration (FRC): in the pulsed approach pursued by Helion Energy,
FRC plasmoids are formed inductively, accelerated, merged, and
compressed to fusion conditions by external coils \cite{Slough2011,Steinhauer2011,Tuszewski1988,Slough1993}. In
such plasmas the pressure is comparable to the magnetic pressure and the
ion gyroradii are comparable to gradient scale lengths, so the coupling
between particles and fields is fundamentally kinetic at the ion scale,
while the phenomena of interest (global stability, flux and energy
transport, compressional heating) unfold over thousands of ion transit and Alfv\'en times. Fully kinetic electromagnetic particle-in-cell (PIC) simulation
must additionally resolve the electron plasma frequency, the Debye
length, and the speed-of-light CFL constraint, placing such
long-timescale studies out of reach. The hybrid PIC model, in which kinetic ions are coupled to an inertialess, charge-neutralizing electron fluid closed by a generalized Ohm's law, removes the electron scales from the problem, so that the resolution requirements are set by the ion dynamics alone \cite{Nielson1976,WinskeHybrid2003}. An explicit Ohm's-law hybrid solver of
this type was implemented in the open-source, GPU-capable PIC code WarpX
\cite{Vay2018} and applied to global stability studies of high-beta fusion
plasmas by Groenewald \emph{et al.}~\cite{Groenewald2023}.

Explicit hybrid schemes, however, face a stiff stability constraint similar to that in Hall MHD. Ohm's law coupled to Faraday's law supports whistler waves,
whose frequency grows quadratically with wavenumber, so that the explicit
time step must satisfy $\dt \propto \Delta x^{2}/(d_i v_A)$, with $d_i$ the
ion inertial length and $v_A$ the Alfv\'en speed \cite{ChaconKnoll2003}: a
factor-of-two grid refinement costs a factor of eight in time steps. The
standard mitigations each carry a penalty. Sub-stepping the field advance
with frozen ion sources multiplies the field-solve cost and introduces
splitting error; density floors regularize the $1/\nel$ factor in Ohm's law
at the expense of ad hoc vacuum physics; and hyper-resistivity damps the
offending grid-scale whistlers but also dissipates physical modes, an
artificial loss that accumulates over long simulations and can degrade
global observables such as the trapped magnetic flux. For simulations that aim at transport timescales, it is these accumulated errors, rather than raw cost, that are often the binding constraint.

Implicit time integration offers a principled alternative and has a
well-developed history in fully kinetic PIC. Energy-conserving fully
implicit PIC methods were established by Markidis and
Lapenta~\cite{MarkidisLapenta2011} and Chen, Chac\'on, and
Barnes~\cite{ChenChaconBarnes2011}, the latter introducing the nonlinear
elimination (``particle-suppressed'') strategy that makes Jacobian-free
Newton--Krylov (JFNK) solution \cite{KnollKeyes2004} of the coupled
field--particle system practical. The $\theta$-implicit variant, with a
time-biasing parameter that interpolates between the energy-conserving
midpoint rule and selectively damped advances, was analyzed in detail by
Angus \emph{et al.}~\cite{Angus2022,Angus2023}, who also adapted the mass
matrices of the energy-conserving semi-implicit (ECSIM) method
\cite{Lapenta2017} to remove all particle operations from the linear stage
of the JFNK solve. For the hybrid model specifically, Stanier, Chac\'on, and Chen \cite{StanierChacon2019} developed a fully implicit, conservative, nonlinear scheme: a JFNK-solved hybrid PIC with implicit-midpoint time advance, exact energy conservation, and ion sub-cycling with orbit averaging, subsequently extended to curvilinear meshes \cite{Stanier2022}.

This paper presents a $\theta$-implicit hybrid PIC algorithm, implemented
in WarpX, that brings these threads together in a production, GPU-capable
code and extends them in several directions. The fields and particles are
advanced as a single coupled nonlinear system: the particles are
nonlinearly eliminated, the magnetic field is eliminated through Faraday's
law, and, because the electric field is algebraic in this model, the
residual reduces to a fixed-point condition on $\vE$ alone. The
time-biasing parameter $\theta \in [1/2, 1]$ is adjustable; at
$\theta = 1/2$ the advance coincides with the implicit midpoint rule of
Ref.~\cite{StanierChacon2019}, while $\theta > 1/2$ provides
selective damping of under-resolved whistler modes. The electron closure
is advanced as a discrete pressure equation inside the nonlinear solve,
so that resistive and hyper-resistive heating enter the electron fluid
through the same discrete work terms that act on the ions and the
magnetic field, and an energy-conserving electron--ion temperature
relaxation couples the electron fluid back to the ion particles. We
derive the discrete energy law of the scheme, identifying the
conservative, dissipative, and exchange channels of the hybrid model and
a single positive-definite numerical damping term that vanishes
identically at $\theta = 1/2$. The mass-matrix acceleration of the linear
stage is adapted to the Ohm's-law Jacobian, in which the deposited ion
current enters through the Hall, pressure, and friction terms rather than
through a displacement current. Externally applied fields derived from a
time-dependent vector potential are supported through an exact
internal--external splitting, enabling coil-driven compression and flux
ramps. The method is verified against electromagnetic wave dispersion in
1D and RZ geometry, cold-beam finite-grid-instability and
energy-conservation benchmarks, and the $m=0$ instability of a Bennett
pinch in RZ geometry, and applied to 3D simulations of FRC plasmas that compare explicit and implicit advances.

The remainder of the paper is organized as follows.
Section~\ref{sec:model} presents the hybrid model, the $\theta$-implicit
discretization, the external-field splitting, and the discrete energy
law. Section~\ref{sec:solver} describes the JFNK solution of the
nonlinear system, the implicit particle advance, and the mass matrices.
Section~\ref{sec:results} presents verification and numerical results,
and Sec.~\ref{sec:performance} the solver performance. A summary and
outlook are given in Sec.~\ref{sec:summary}, and the detailed derivation
of the discrete energy law in Appendix~\ref{app:energy}.

\section{\label{sec:model}The hybrid plasma model and $\theta$-implicit discretization}

\subsection{\label{sec:model:ohm}Governing equations and generalized Ohm's law}

In the hybrid model considered here, each ion species $s$ is represented by
kinetic macro-particles while the electrons are treated as an inertialess,
charge-neutralizing fluid \cite{Nielson1976,WinskeHybrid2003}. The
non-relativistic equations of motion for macro-particle $p$ are
\begin{equation}
\frac{d\vx_p}{dt} = \vv_p, \qquad
\frac{d\vv_p}{dt} = \frac{q_p}{m_p}
 \left[ \vE(\vx_p) + \vv_p\times\vB(\vx_p) \right],
\label{eq:ion_eom}
\end{equation}
where $m_p$ and $q_p$ are the macro-particle mass and charge. The model is
valid for dynamics slow compared to the electron plasma and cyclotron
frequencies and for phase velocities small compared to the speed of light. In
this ordering two approximations are made together: quasineutrality,
$e\nel = \sum_s q_s n_s$, and the neglect of the displacement current in
Amp\`ere's law \cite{Nielson1976},
\begin{equation}
\mu_0 \vJ = \curl\vB,
\label{eq:ampere}
\end{equation}
so that the total current density is available directly from the magnetic
field. The electron current density then follows by subtraction,
\begin{equation}
\vJe = \vJ - \vJi - \vJext,
\qquad \vJi = \sum_{s} \vJ_s,
\label{eq:je}
\end{equation}
where $\vJi$ is deposited from the ion macro-particles and $\vJext$ is any
prescribed external current not represented by particles. The electric field is obtained from the massless-electron momentum equation,
which takes the form of a generalized Ohm's law
\cite{WinskeHybrid2003,Groenewald2023,Belyaev2024},
\begin{equation}
\vE = \frac{1}{e\nel}\left( \vJe\times\vB - \grad\pe \right)
 + \sum_s f_s\,\eta_s\, e\nel \left( \bm{V}_s - \vVe \right)
 - \etah\grad^2\vJ,
\label{eq:ohm}
\end{equation}
where $\pe$ is the scalar electron pressure and $\etah(\rho,|\vB|)$ a
user-specified hyper-resistivity. The second term is the collisional
friction of the electrons against each ion species $s$
\cite{Belyaev2024}: $\eta_s$ is the corresponding resistivity (a
user-specified function of the local densities, electron temperature, currents, magnetic field strength,
and time), $f_s = \rho_s/\sum_t \rho_t$ is the charge-density fraction with
$\rho_s = q_s n_s$, $\bm{V}_s = \vJ_s/\rho_s$ is the species drift
velocity, and $\vVe = -\vJe/(e\nel)$ is the electron fluid velocity. This
permits, for example, a temperature-dependent Spitzer drag against one
species on top of a constant background resistivity. For a resistivity
common to all species, quasineutrality gives
$\sum_s f_s\, e\nel(\bm{V}_s - \vVe) = \vJ$ and the friction reduces to the
familiar $\eta\vJ$. It is convenient to collect the frictional and
hyper-resistive terms into the dissipative field
\begin{equation}
\vD = \sum_s f_s\,\eta_s\, e\nel \left( \bm{V}_s - \vVe \right)
 - \etah\grad^2\vJ ,
\label{eq:dfield}
\end{equation}
which plays a central role in the discretization below.

The magnetic field evolves according to Faraday's law,
\begin{equation}
\frac{\partial \vB}{\partial t} = -\curl\vE.
\label{eq:faraday}
\end{equation}
The system is closed by the evolution of the scalar electron pressure,
\begin{equation}
\frac{\partial\pe}{\partial t} + \vVe\cdot\grad\pe
 + \gamma\,\pe\,\diver\vVe = (\gamma-1)\,S_e,
\label{eq:electron_energy}
\end{equation}
the internal-energy equation of the electron fluid with adiabatic index
$\gamma$; thermal conduction is neglected in the present implementation.
The source is the dissipative heating together with electron--ion
temperature relaxation,
\begin{equation}
S_e = \vD\cdot\vJ - Q_{ei},
\qquad
Q_{ei} = \sum_s 3\, n_s k_B\, \nu_{ei}\left( \Te - T_{i,s} \right),
\label{eq:qei}
\end{equation}
where $\Te = \pe/(\nel k_B)$, $T_{i,s}$ is the local ion temperature of
species $s$, and $\nu_{ei}$ is a user-specified relaxation rate
\cite{Belyaev2024}. The heating deposited in the electrons by the
frictional and hyper-resistive terms of \cref{eq:ohm} is the work of the
dissipative field, $\vD\cdot\vJ$, and the exchange $Q_{ei}$ is paired with
matching heating of the ion macro-particles
(Sec.~\ref{sec:model:theta}), so it moves energy between the species.

For $S_e = 0$, \cref{eq:electron_energy} conserves the electron entropy
function $\pe\,\nel^{-\gamma}$ along the electron-fluid characteristics.
For the uniform-entropy initial condition set by a reference state
$(n_0, T_{e0})$, the pressure then remains on the global adiabat
\begin{equation}
\pe = n_0 k_B T_{e0} \left( \frac{\nel}{n_0} \right)^{\gamma},
\label{eq:closure}
\end{equation}
the familiar polytropic closure, with $\gamma = 1$ the isothermal and
$\gamma = 5/3$ the adiabatic limit; the general case departs from
\cref{eq:closure} exactly where dissipation or exchange generates entropy
locally. The pressure equation is integrated inside the nonlinear solve
(Sec.~\ref{sec:model:theta}), and the discrete form of its work terms is
essential to the energy law of Sec.~\ref{sec:model:energy}.

The stiffest normal mode supported by \cref{eq:ohm,eq:faraday} is the
whistler wave. For parallel propagation in a uniform plasma the dispersion
relation is \cite{ChaconKnoll2003}
\begin{equation}
\omega^2 = v_A^2 k^2 \left[ 1 + \frac{d_i^2 k^2}{2}
 + d_i k \sqrt{1 + \frac{d_i^2 k^2}{4}} \,\right],
\label{eq:whistler_disp}
\end{equation}
where $v_A = B/\sqrt{\mu_0 n_i m_i}$ is the Alfv\'en speed and
$d_i = c/\omega_{pi}$ the ion inertial length. In the short-wavelength limit
$k d_i \gg 1$ the frequency grows quadratically, $\omega \simeq v_A d_i k^2$,
so that explicit stability at the grid Nyquist wavenumber
$k_\mathrm{max} = \pi/\Delta x$ requires
\begin{equation}
\dt \le \frac{1}{\pi}\frac{\Delta x^2}{d_i\, v_A}.
\label{eq:whistler_cfl}
\end{equation}
The quadratic dependence on $\Delta x$ makes grid refinement rapidly
prohibitive for explicit schemes and motivates the implicit time
discretization described next.

\subsection{\label{sec:model:theta}The $\theta$-implicit time discretization}

The system is advanced from $t^n$ to $t^{n+1} = t^n + \dt$ by solving for all
quantities at the intermediate time $t^{n+\theta}$, where for an arbitrary
quantity $h$,
\begin{equation}
\ntheta{h} = (1-\theta)\, h^{n} + \theta\, h^{n+1},
\qquad \tfrac{1}{2} \le \theta \le 1 .
\label{eq:theta_def}
\end{equation}
The discrete equations of motion for the macro-particles are the
Crank--Nicolson forms
\begin{equation}
\frac{\vx_p^{n+1}-\vx_p^{n}}{\dt} = \bar{\vv}_p,
\qquad
\frac{\vv_p^{n+1}-\vv_p^{n}}{\dt} =
 \frac{q_p}{m_p}\left( \vEstar_{p}
 + \bar{\vv}_p\times\ntheta{\vB}_p \right),
\label{eq:particle_cn}
\end{equation}
where $\bar{\vv}_p \equiv \nhalf{\vv}_p = (\vv_p^{n+1}+\vv_p^{n})/2$, the
fields are gathered at the time-centered position
$\bar{\vx}_p = (\vx_p^{n+1}+\vx_p^{n})/2$, and $\vEstar$ is the push field
defined in \cref{eq:estar} below. The time-centered ion current
$\nhalf{\vJi}$ and charge density $\nhalf{\rho}$ are deposited from
$(\bar{\vx}_p, \bar{\vv}_p)$ using the same shape function employed to gather
the electric field to the particles, $S^E_{gp} = S^J_{gp}$; this consistency
condition is required for the energy analysis of
Sec.~\ref{sec:model:energy} \cite{MarkidisLapenta2011,Angus2023}. The
magnetic field is advanced with the discrete Faraday law,
\begin{equation}
\ntheta{\vB} = \vB^{n} - \theta\,\dt\,\curl\ntheta{\vE},
\label{eq:faraday_discrete}
\end{equation}
with $\vB^{n+1} = \vB^{n} - \dt\,\curl\ntheta{\vE}$.

The electron pressure is advanced with the discrete form of
\cref{eq:electron_energy}. Within every residual evaluation, after the
particle advance and deposition, $\ntheta{\pe}$ is obtained from
\begin{equation}
\frac{\ntheta{\pe} - \pe^{n}}{\theta\,\dt}
 = -\gamma\,\diver\!\left(\ntheta{\pe}\,\vVe\right)
 + (\gamma-1)\left[\, \ntheta{\vE}\cdot\vJe
 + \vD\cdot\nhalf{\vJi} \,\right],
\label{eq:pe_advance}
\end{equation}
with $\vVe$ and $\vJe$ evaluated from the iterate's moments. Two discrete
choices give the scheme its conservation property: the transport term is
written in conservative flux form, so its sum over the grid vanishes
identically, and the work terms in the bracket are evaluated with the
same discrete fields, at the same mesh locations, as the ion push
\cref{eq:particle_cn} and the Faraday update \cref{eq:faraday_discrete}.
The update is $\theta$-centered in $\pe$ by a short fixed-point iteration
and a positivity floor is enforced.

Equations~(\ref{eq:particle_cn})--(\ref{eq:pe_advance}), together with
Ohm's law evaluated at $t^{n+\theta}$, constitute a coupled nonlinear system
for the fields and particles,
$\bm{F}(\bm{X}) = [\bm{F}_f(\bm{X}_f,\bm{X}_p);\,
\bm{F}_p(\bm{X}_f,\bm{X}_p)] = \bm{0}$. Its dimensionality is reduced in two
steps. First, the particle quantities are nonlinearly eliminated: for given
fields, \cref{eq:particle_cn} is solved for each particle independently by
Picard iteration, so the particle unknowns become implicit functions of the
field unknowns \cite{ChenChaconBarnes2011,Angus2023,StanierChacon2019}.
Second, the magnetic field is eliminated through
\cref{eq:faraday_discrete}. Because $\vE$ carries no evolution equation of
its own, the remaining residual is a
fixed-point condition on the electric field alone,
\begin{equation}
\bm{F}\!\left(\ntheta{\vE}\right) = \ntheta{\vE}
 - \vE_\mathrm{Ohm}\!\left( \ntheta{\vB}(\ntheta{\vE}),\,
   \nhalf{\vJi},\, \nhalf{\rho},\, \ntheta{\pe} \right) = \bm{0},
\label{eq:residual}
\end{equation}
where each evaluation of $\vE_\mathrm{Ohm}$ entails a self-consistent
particle re-advance, deposition, and pressure advance. The solution of
\cref{eq:residual} with a Jacobian-free Newton--Krylov method is described
in Sec.~\ref{sec:solver}.

The field gathered to the particles in \cref{eq:particle_cn} is not the
full Ohm field: the terms collected in $\vD$ represent the collisional drag
and viscosity of the electrons, their reaction on the ions is not otherwise
contained in the model, and gathering them to the ions would spuriously
accelerate them \cite{StanierChacon2019}. The particles are therefore
pushed with
\begin{equation}
\vEstar = \ntheta{\vE} - \vD,
\label{eq:estar}
\end{equation}
where $\ntheta{\vE}$ is the Newton iterate itself. Pushing with the
iterate, rather than with a field recomputed from lagged sources, makes the
residual a differentiable function of the unknown through the particle
response; in particular the electrostatic limit ($\vB = 0$), where a
recomputed push field would not depend on $\ntheta{\vE}$ at all, is well
posed. For the same reason, the density entering Ohm's law and the pressure
advance is the one deposited from the current particle iterate. The
discrete $\vD$ is formed once per residual evaluation as the difference of
two Ohm's-law evaluations from the same state, with and without the
dissipative terms: the ideal, Hall, and pressure contributions cancel
exactly, and $\vD$ inherits the stencils, interpolations, and floors of the
residual by construction. The full Ohm field is used in
\cref{eq:faraday_discrete} and in the residual \cref{eq:residual}. (The
pressure gradient is curl-free on the continuum level and does not affect
the Faraday update, but it is retained in the residual for consistency of
the Newton linearization.) The role of $\vEstar$ and $\vD$ in the energy
budget is made explicit in Sec.~\ref{sec:model:energy}.

After the nonlinear iteration converges, all field quantities, including
the electron pressure, are extrapolated from $t^{n+\theta}$ to $t^{n+1}$,
\begin{equation}
h^{n+1} = \frac{1}{\theta}\,\ntheta{h}
 + \left(1 - \frac{1}{\theta}\right) h^{n},
\label{eq:extrap}
\end{equation}
and the particles are advanced from $t^{n+1/2}$ to $t^{n+1}$. For
$\theta = 1/2$ the scheme is the implicit midpoint rule and is second-order
accurate in time; this special case coincides with the time advance of the
conservative implicit hybrid scheme of Stanier
\emph{et al.}~\cite{StanierChacon2019,Stanier2022}. For $\theta > 1/2$ the
scheme is first-order accurate and introduces selective numerical damping of
modes with frequencies approaching $1/\dt$, quantified in
Sec.~\ref{sec:model:energy}.

The exchange $Q_{ei}$ of \cref{eq:qei} is applied once per step, after the
fields and particles reach $t^{n+1}$: the conjugate ion heating must
modify the ion macro-particles and cannot enter the residual
\cref{eq:residual}. The sink on the electron pressure and the
thermal-velocity kick on the ion macro-particles of each species are
applied with equal magnitude, so the exchange moves energy between the two
species without creating or destroying it.

\subsection{\label{sec:model:extfields}External fields via vector-potential splitting}

Many applications of interest are driven by prescribed external circuits, such as compression coils, bias-field ramps, or flux conservers, whose fields can
greatly exceed the plasma-generated fields. These are supported through an
exact splitting of the fields into internal (plasma-generated) and external
(prescribed) parts,
\begin{equation}
\vB = \vB_\mathrm{int} + \vB_\mathrm{ext}, \qquad
\vE = \vE_\mathrm{int} + \vE_\mathrm{ext},
\label{eq:ext_split}
\end{equation}
with the external pair derived from a user-specified vector potential
$\vA(\vx)$ modulated by a temporal waveform $f(t)$,
\begin{equation}
\vB_\mathrm{ext} = f(t)\,\curl\vA, \qquad
\vE_\mathrm{ext} = -\vA\,\frac{df}{dt},
\label{eq:ext_fields}
\end{equation}
where $df/dt$ is evaluated with a time-centered finite difference consistent
with the $\theta$-scheme. By construction the external pair satisfies
Faraday's law identically,
$\partial\vB_\mathrm{ext}/\partial t = -\curl\vE_\mathrm{ext}$, so the
splitting is exact: the discrete Faraday advance
\cref{eq:faraday_discrete} applies to $\vB_\mathrm{int}$ unchanged and no
splitting error is introduced.

The nonlinear solver operates on the internal fields only. Within each
residual evaluation, the plasma current is computed from the internal field,
$\mu_0\vJ = \curl\vB_\mathrm{int}$, consistent with attributing
$\curl\vB_\mathrm{ext}/\mu_0$ to the external coil current $\vJext$ excluded
in \cref{eq:je}; the Hall term in Ohm's law is evaluated with the total
magnetic field; and the particles are pushed with the total fields
$(\vEstar+\vE_\mathrm{ext},\, \vB_\mathrm{int}+\vB_\mathrm{ext})$.
In the vacuum region exterior to the plasma the Ohm field vanishes and the
total field reduces to the prescribed external drive. The external drive
enters the energy budget of Sec.~\ref{sec:model:energy} as a physical input
power through the work done on the particles and the Poynting flux of the
external fields.

\subsection{\label{sec:model:energy}Discrete energy law}

We now state the discrete energy budget of the scheme, following the
approach used for the fully electromagnetic $\theta$-PIC method
\cite{MarkidisLapenta2011,Angus2022,Angus2023} but adapted to the hybrid
model, in which the electric field carries no energy degree of freedom,
Ohm's law contains dissipative terms, and the electron internal energy is
tracked by \cref{eq:pe_advance}. For simplicity we consider periodic
boundaries and no external drive; surface contributions and the external
input power are restored in~\ref{app:energy}. The energies at step
$n$ are
\begin{equation}
W_B^{n} = \sum_g \frac{|\vB_g^{n}|^2}{2\mu_0}\,\DV,
\quad
W_i^{n} = \sum_p \frac{m_p}{2}\,|\vv_p^{n}|^2,
\quad
W_e^{n} = \sum_g \frac{P_{e,g}^{n}}{\gamma-1}\,\DV .
\label{eq:energy_defs}
\end{equation}

Three per-channel identities hold, derived in~\ref{app:energy}.
For the particles, the Crank--Nicolson update \cref{eq:particle_cn} gives
$W_i^{n+1}-W_i^{n} = \sum_p m_p \bar{\vv}_p\cdot(\vv_p^{n+1}-\vv_p^{n})$;
the magnetic force does no work exactly, and provided the gather and
deposit shape functions coincide ($S^E_{gp}=S^J_{gp}$) and both are
evaluated at the time-centered particle positions, the sum over particles
collapses onto the grid \cite{MarkidisLapenta2011,Angus2023}:
\begin{equation}
\frac{W_i^{n+1}-W_i^{n}}{\dt} =
 \sum_g {\nhalf{\vJi}}_{g}\cdot\vEstar_{g}\,\DV .
\label{eq:wi_budget}
\end{equation}
On staggered grids this condition requires the momentum-conserving
(non-Galerkin) field interpolation; on the collocated grid it holds by
default. For the magnetic field, multiplying the discrete Faraday law by
the time-averaged field $\bar{\vB} = (\vB^{n+1}+\vB^{n})/2$, using
$\bar{\vB} = \ntheta{\vB} - \phi\,(\vB^{n+1}-\vB^{n})$ with
$\phi \equiv \theta - 1/2$, and applying the mutual adjointness of the
discrete curl operators together with
$\mu_0\ntheta{\vJ} = \curl\ntheta{\vB}$,
\begin{align}
\frac{W_B^{n+1}-W_B^{n}}{\dt} ={}&
 -\sum_g \ntheta{\vJ}_g\cdot\ntheta{\vE}_g\,\DV \nonumber\\
& - \frac{\phi}{\mu_0\,\dt}
   \sum_g \left| \vB_g^{n+1}-\vB_g^{n} \right|^2 \DV,
\label{eq:wb_budget}
\end{align}
where $\ntheta{\vE}$ is the full Ohm field used in the Faraday update. The
second term is negative definite for $\phi > 0$ and vanishes identically
for $\theta = 1/2$; it is the hybrid counterpart of the numerical damping
term of the electromagnetic $\theta$-PIC scheme \cite{Angus2022}, with
only the magnetic contribution present since $\vE$ is not a dynamical
variable of the model. For the electrons, summing \cref{eq:pe_advance}
over the grid, the flux term telescopes and, with \cref{eq:extrap},
\begin{equation}
\frac{W_e^{n+1}-W_e^{n}}{\dt} =
 \sum_g \left( \ntheta{\vE}\cdot\vJe
 + \vD\cdot\nhalf{\vJi} \right)_g \DV .
\label{eq:we_budget}
\end{equation}

Adding \cref{eq:wi_budget,eq:wb_budget,eq:we_budget} and using
$\vEstar = \ntheta{\vE} - \vD$ and $\vJe = \ntheta{\vJ} - \nhalf{\vJi}$,
every work term cancels in pairs and the total energy obeys
\begin{equation}
\frac{(W_i{+}W_B{+}W_e)^{n+1}-(W_i{+}W_B{+}W_e)^{n}}{\dt} =
 - \frac{\phi}{\mu_0\,\dt}
   \sum_g \left| \vB_g^{n+1}-\vB_g^{n} \right|^2 \DV .
\label{eq:energy_law}
\end{equation}
The cancellation does not reference the contents of Ohm's law: any
resistivity model, hyper-resistivity, or density floor entering
$\ntheta{\vE}$ and $\vD$ is covered, because the electron budget
\cref{eq:we_budget} is constructed to receive the exact complement of the
discrete ion work and magnetic energy change. The individual channels
retain their physical meaning: the Hall term transfers energy between the field and the ions and does no
net work on the electrons; the dissipative work
$\vD\cdot\vJ$ appears in the electron internal energy as Joule and
hyper-resistive heating; and the pressure term is the reversible
compressional exchange.

The conditions for \cref{eq:energy_law} to hold exactly are: (i)~the
gather/deposit consistency $S^E_{gp}=S^J_{gp}$ underlying
\cref{eq:wi_budget}; (ii)~mutually adjoint curl operators in
\cref{eq:wb_budget}; and (iii)~agreement between the discrete products of
\cref{eq:we_budget} and those of the ion and Faraday updates, together
with the conservative flux form of the transport term in
\cref{eq:pe_advance}. All three are satisfied in Cartesian geometry, on the collocated and the
Yee-staggered mesh alike, and the total energy is then conserved at
$\theta = 1/2$ independent of the time step. Because the cancellation is
algebraic in the accepted iterate, the conservation error is set by
floating-point roundoff rather than by the tolerance of the nonlinear
solve: an incompletely converged solve appears as an error in the fields,
not in the energy balance. The measured conservation is reported in
Sec.~\ref{sec:results}. 
For $\theta > 1/2$ the $\phi$ term is the only
numerical dissipation; it scales with the per-step field increment and
therefore acts selectively on under-resolved (whistler-scale)
fluctuations, with its cumulative effect examined in
Sec.~\ref{sec:results:frc}.

\section{\label{sec:solver}The nonlinear solver}

\subsection{\label{sec:solver:jfnk}Jacobian-free Newton--Krylov formulation}

The residual \cref{eq:residual} defines a nonlinear system
$\bm{F}(\bm{U}) = \bm{0}$ for the field unknown
$\bm{U} \equiv \ntheta{\vE}$. It is solved with a Jacobian-free
Newton--Krylov (JFNK) method \cite{KnollKeyes2004}. At each Newton
iteration $k$, the correction $\delta\bm{U}$ is obtained from the linear
system
\begin{equation}
\bm{F}'\!\left(\bm{U}^{k}\right)\delta\bm{U} = -\bm{F}\!\left(\bm{U}^{k}\right),
\qquad
\bm{U}^{k+1} = \bm{U}^{k} + \delta\bm{U},
\label{eq:newton}
\end{equation}
which is solved iteratively with GMRES, a Krylov method that requires only
the action of the Jacobian on a Krylov vector $\bm{v}$, evaluated matrix-free by a
finite-difference directional derivative,
\begin{equation}
\bm{F}'\!\left(\bm{U}^{k}\right)\bm{v} \approx
 \frac{\bm{F}\!\left(\bm{U}^{k}+\epsilon\,\bm{v}\right)
     - \bm{F}\!\left(\bm{U}^{k}\right)}{\epsilon},
\qquad
\epsilon = \epsilon_0\,
 \frac{\sqrt{1+\lVert\bm{U}^{k}\rVert}}{\lVert\bm{v}\rVert},
\label{eq:jfnk_fd}
\end{equation}
with $\epsilon_0 \sim 10^{-6}$ chosen small enough for accuracy of the
linearization and large enough to avoid floating-point roundoff
\cite{KnollKeyes2004}. The Newton iteration is declared converged when
$\lVert\bm{F}\rVert$ falls below
$\max\!\left( r_\mathrm{tol}\lVert\bm{F}^{0}\rVert,\, a_\mathrm{tol}\right)$.

The structure of \cref{eq:residual} makes each residual evaluation
expensive: it entails the construction of $\vD$ and $\vEstar$, a
self-consistent implicit particle advance
(Sec.~\ref{sec:solver:particles}), deposition of $\nhalf{\vJi}$ and
$\nhalf{\rho}$, the Faraday update of $\ntheta{\vB}$, the pressure advance
\cref{eq:pe_advance}, and the evaluation of Ohm's law. In the baseline
JFNK method this cost is incurred at every Newton \emph{and} every GMRES
iteration. This is the particle-suppressed JFNK strategy
\cite{ChenChaconBarnes2011,Angus2023}, in which the particle degrees of
freedom are nonlinearly eliminated rather than carried as unknowns; the
mass-matrix acceleration of Sec.~\ref{sec:solver:massmatrices} removes the
particle operations from the GMRES iterations.

The implementation also provides a Picard (fixed-point) iteration on
$\ntheta{\vE}$ as a simpler alternative nonlinear solver, and optional
PETSc-based Newton (SNES) and Krylov (KSP) backends; all results in this
paper use the native JFNK solver.
%
\begin{figure*}[htb]
\includegraphics[width=\textwidth]{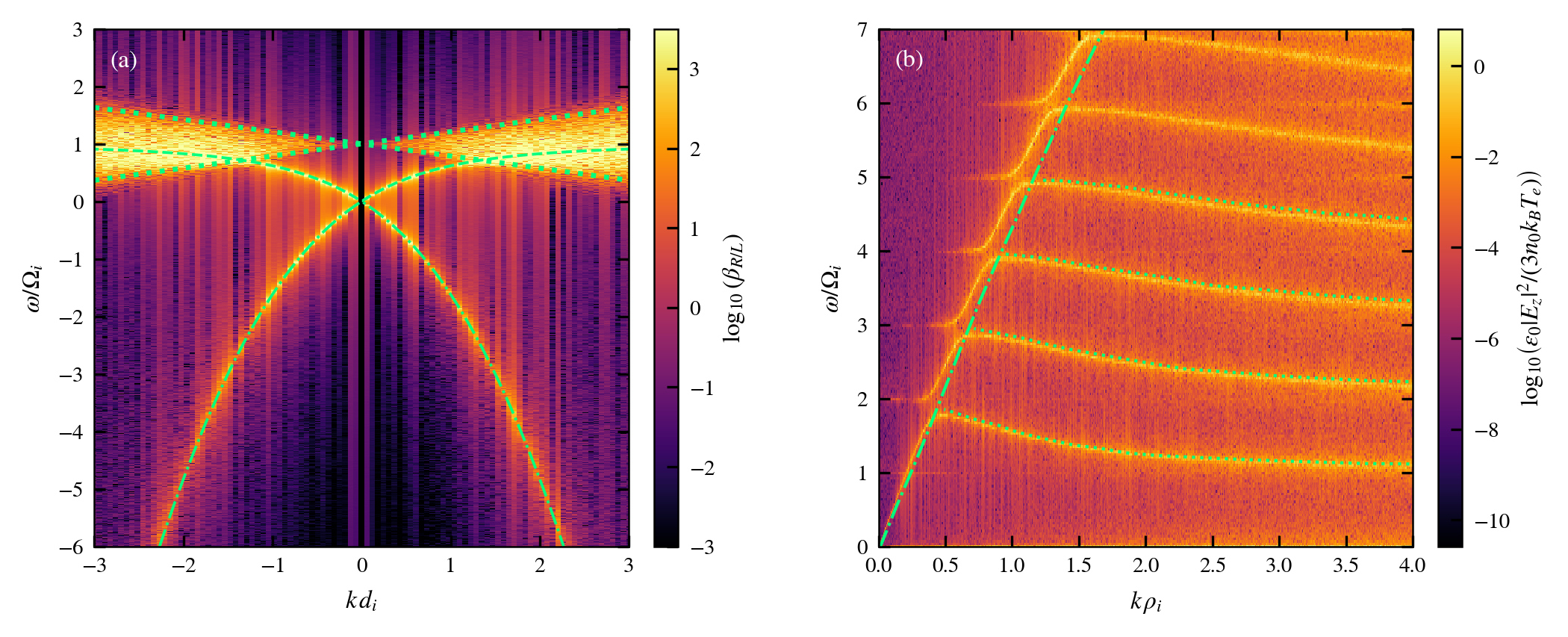}
\caption{\label{fig:dispersion}Dispersion verification of the implicit
solver ($\theta = 1/2$). (a)~Parallel propagation: $\omega$--$k$ spectrum
of the circularly polarized fluctuation $B_{R/L} = B_x \pm i B_y$, with
the cold-plasma L and R branches,
$(k d_i)^2 = (\omega/\Omega_i)^2/(1 \pm \omega/\Omega_i)$ (dashed and
dash-dotted), and the thermally broadened ion-cyclotron lines
$\omega = \Omega_i \pm 3 v_{ti} k$ (dotted); $\dt =
5\times10^{-4}\,\tau_{ci}$, spectrum accumulated over $222\,\tau_{ci}$.
(b)~Perpendicular propagation: $\omega$--$k$ spectrum of $E_z$, with the
fast magnetosonic/extraordinary branch (dash-dotted) and the ion
Bernstein harmonics (dotted) of Ref.~\cite{Munoz2018}; $\dt =
1\times10^{-4}\,\tau_{ci}$, spectrum accumulated over $28\,\tau_{ci}$.
Both runs use the small reference resistivity
$\eta = 10^{-7}~\Omega$\,m.}
\end{figure*}
\begin{figure*}[htb]
\includegraphics[width=\textwidth]{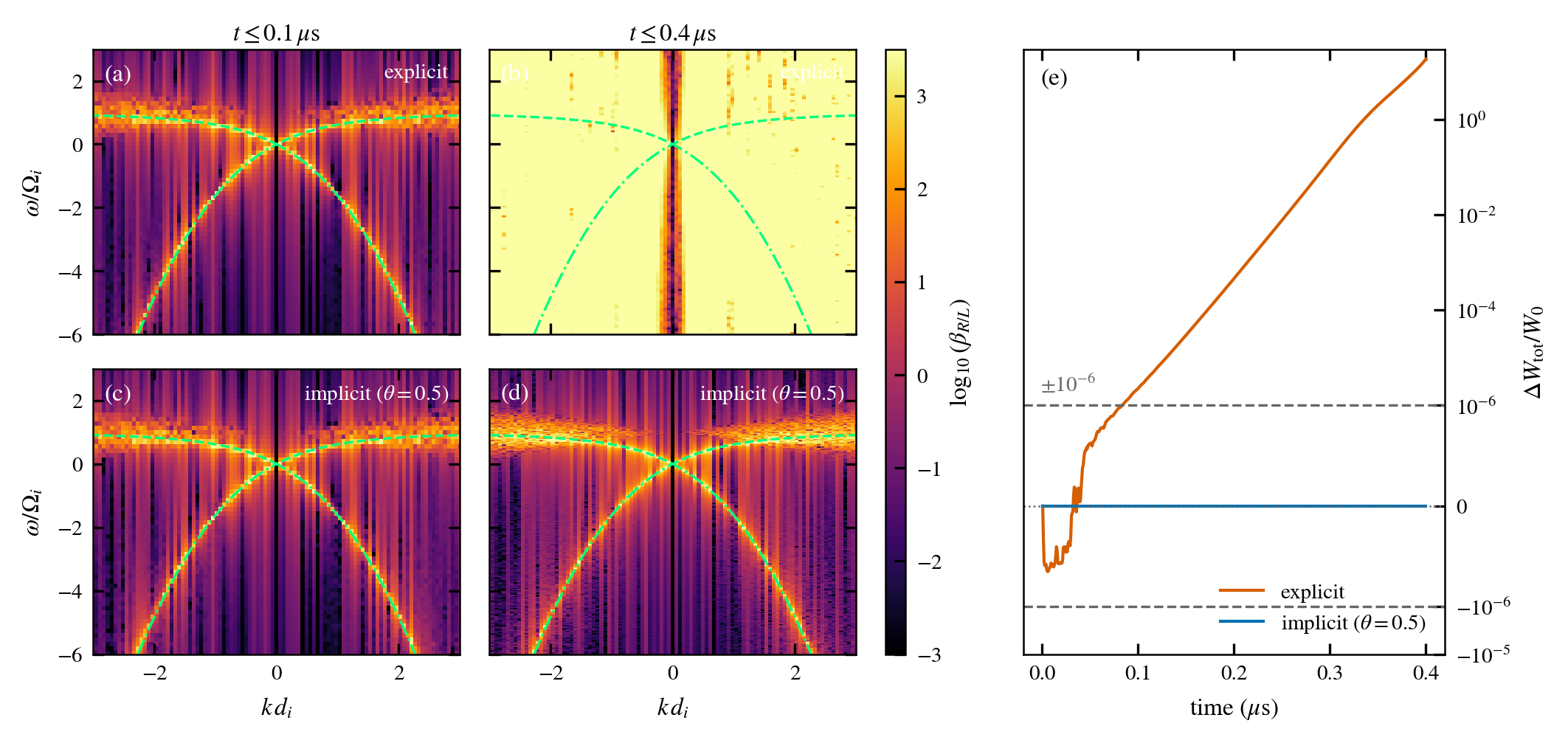}
\caption{\label{fig:expimp}Explicit versus implicit advance for the
parallel EM-mode configuration with $\eta = \etah = 0$. (a,b)~Spectra of
$B_{R/L}$ from the explicit solver ($\dt = 6.25\times10^{-4}\,\tau_{ci}$,
40 field sub-steps per particle step), computed over cumulative windows
$t \le 0.1~\mu$s and $t \le 0.4~\mu$s. (c,d)~The same windows for the
implicit solver ($\theta = 1/2$, $\dt = 5\times10^{-4}\,\tau_{ci}$, no
sub-stepping). Dispersion overlays as in Fig.~\ref{fig:dispersion}(a).
(e)~Relative change in the total energy $W_i + W_B + W_e$: the explicit
error grows secularly, crossing $10^{-6}$ near $t = 0.09~\mu$s and
reaching $\Delta W_\mathrm{tot}/W_0 = 19$ at $0.4~\mu$s, while the
implicit error remains below $10^{-13}$ throughout. Dashed lines mark
$\pm 10^{-6}$, the Newton relative tolerance, shown for reference.}
\end{figure*}
\begin{figure*}[htbp]
\includegraphics[width=\textwidth]{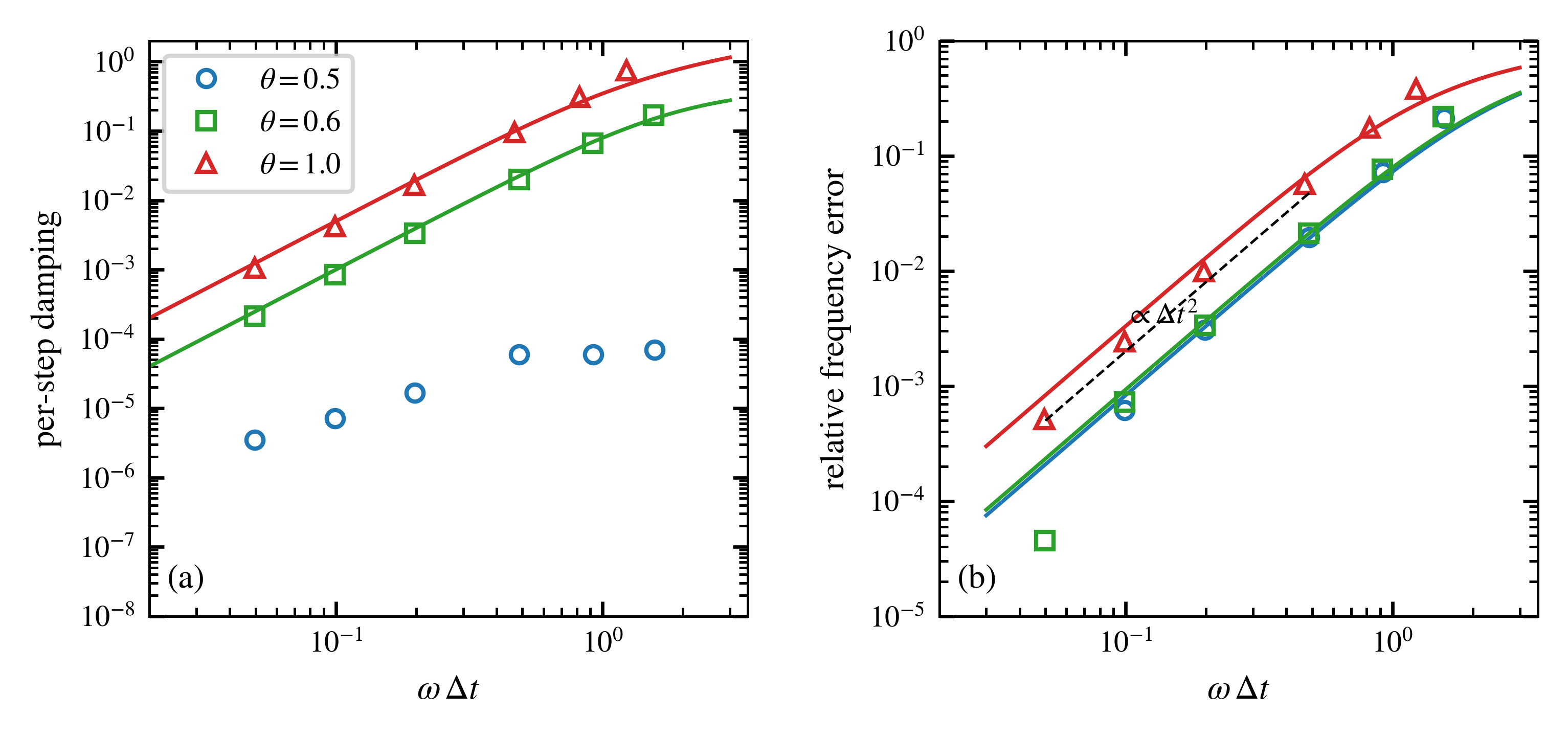}
\caption{\label{fig:theta}Single-whistler-mode response of the
$\theta$ scheme ($k d_i = 2.0$, $\omega \simeq 4.7\,\Omega_i$) as a
function of $\omega\dt$, for three values of the time-biasing parameter
(symbols), against the analytic amplification factor
$G = [1 - (1-\theta)i\omega\dt]/[1 + \theta i\omega\dt]$ (lines).
(a)~Per-step damping $-\ln|G|$, with $|G| = |a^{n+1}|/|a^n|$ the
one-step amplitude ratio. At $\theta = 1/2$ the numerical damping
vanishes and the measured points sit at the weak physical damping of
the mode; for $\theta > 1/2$ the damping follows the predicted
$(\theta - 1/2)(\omega\dt)^2$ ramp. (b)~Relative error of the measured
mode frequency: second order in $\dt$ for all $\theta$, with the
coefficient $[\theta^3 + (1-\theta)^3]/3$ of the scheme's phase error
(equal to $1/12$ at $\theta = 1/2$).}
\end{figure*}
\subsection{\label{sec:solver:particles}Implicit particle advance}

Each residual evaluation must solve the Crank--Nicolson equations of motion
\cref{eq:particle_cn} for every macro-particle, given the current field
iterate. Because the particles are mutually independent for fixed fields,
this is done particle-by-particle with a Picard iteration
\cite{ChenChaconBarnes2011,Angus2023}. With
$\alpha \equiv q_p\dt/(2m_p)$ and
$\bm{b} \equiv \alpha\,\ntheta{\vB}_p$, the velocity equation in
\cref{eq:particle_cn} is linear in $\bar{\vv}_p$ for fields frozen at the
particle position and admits the closed-form (Boris-like) solution
\begin{equation}
\bar{\vv}_p = \frac{ \vv^{\,\prime} + \vv^{\,\prime}\times\bm{b}
 + \left(\vv^{\,\prime}\cdot\bm{b}\right)\bm{b} }{1 + b^2},
\qquad
\vv^{\,\prime} = \vv_p^{n} + \alpha\,\vEstar_{p} ,
\label{eq:picard_vel}
\end{equation}
where the fields are gathered at the time-centered position
$\bar{\vx}_p$. Iteration $\ell$ gathers the fields at
$\bar{\vx}_p^{(\ell)}$, evaluates \cref{eq:picard_vel}, and re-advances the
position, $\bar{\vx}_p^{(\ell+1)} = \vx_p^{n} + (\dt/2)\,\bar{\vv}_p^{(\ell)}$,
until the position update falls below a prescribed tolerance. As discussed
in Sec.~\ref{sec:model:theta}, the gathered electric field is the
dissipation-free push field $\vEstar$ of \cref{eq:estar}. In practice the
vast majority of particles converge in two to three iterations.

A small population of particles, those with under-resolved gyromotion or displacements spanning multiple cells at large $\dt$, may fail to converge
within the iteration limit. These are flagged, and their trajectories are
integrated by sub-orbits: the step $\dt$ is subdivided per particle into
sub-intervals, each advanced with the same Picard update, with the
sub-orbit count increased until convergence. The flag persists for the
remainder of the time step, so a flagged particle is integrated
consistently in every subsequent residual evaluation, and is reset at the
start of the next step. This per-particle adaptivity confines the cost of
stiff trajectories to the particles that require it, in the spirit of the
orbit sub-cycling employed in other implicit PIC schemes
\cite{ChenChaconBarnes2011,StanierChacon2019}.

\subsection{\label{sec:solver:massmatrices}Mass matrices}

The dominant cost of the baseline algorithm is the particle re-advance and
current deposition inside every GMRES iteration. Following the
energy-conserving semi-implicit (ECSIM) formalism \cite{Lapenta2017} and
its adaptation to the fully implicit setting \cite{Angus2023}, this cost is
removed by linearizing the deposited ion current with respect to the
electric field. If the particle positions $\bar{\vx}_p$ and the magnetic
field in the Lorentz force are held fixed, \cref{eq:picard_vel} makes
$\bar{\vv}_p$ a linear function of the gathered electric field, and the
deposited current becomes an affine function of the grid field,
\begin{equation}
\vJ_{i,g}^{\,n+1/2} = \hat{\vJ}_g
 + \sum_{g'} \mathbb{M}_{gg'}\cdot\ntheta{\vE}_{g'},
\label{eq:massmatrix}
\end{equation}
where $\hat{\vJ}_g$ collects the field-independent part and the mass
matrices $\mathbb{M}_{gg'}$ are banded tensor coefficients with a stencil
set by the particle shape functions \cite{Lapenta2017,Angus2023}. Both are
deposited from the particles once per Newton iteration. During the GMRES
iterations the current perturbation entering the Jacobian action
\cref{eq:jfnk_fd} is then evaluated from \cref{eq:massmatrix} as a purely
grid-based operation, and no particle work is performed. In the linear
stage the charge density is not frozen: it is rebuilt from the mass-matrix
current response through discrete charge continuity,
$\delta\rho = -(\dt/2)\,\diver\delta\vJ_i$, which retains the density
response of the pressure and $1/\nel$ terms in the Jacobian. Particles
flagged for sub-orbit integration (Sec.~\ref{sec:solver:particles}) are
excluded from \cref{eq:massmatrix}; their current is deposited directly
and held as part of $\hat{\vJ}_g$, a current-splitting arrangement
analogous to that suggested in Ref.~\cite{Angus2023}.

Two properties of this construction are worth emphasizing. First, it
modifies only the Jacobian used to compute the Newton correction, while the nonlinear residual is always
evaluated with the full self-consistent particle advance; the converged
solution is therefore unchanged \cite{Angus2023}. Second, in the hybrid
model the mass matrices enter the Jacobian through Ohm's law rather than
through a displacement-current term: a perturbation $\delta\vE$ changes
$\nhalf{\vJi}$ via \cref{eq:massmatrix}, and thereby the electron current
$\vJe = \vJ - \vJi - \vJext$, the Hall and pressure terms of
\cref{eq:ohm}, and the friction. The mass-matrix Jacobian is available in all supported geometries, and is used for the three-dimensional simulations of Sec.~\ref{sec:results:frc}. The resulting
performance gains are quantified in Sec.~\ref{sec:performance}.

\section{\label{sec:results}Verification and numerical results}

\subsection{\label{sec:results:1dmodes}Electromagnetic modes in 1D}

The first verification exercises the linear physics of the model: the
propagation of electromagnetic modes parallel and perpendicular to a
background magnetic field in a uniform thermal plasma. The two
configurations reproduce the verification tests of the explicit
Ohm's-law solver in WarpX \cite{Groenewald2023}, which in turn follow
Sections~4.2 and 4.3 of Mu\~noz \emph{et al.}~\cite{Munoz2018}, allowing
a direct comparison between the explicit and implicit advances on an
established benchmark. Both use a periodic one-dimensional domain with
$\Delta z = 0.1\,d_i$, 1024 particles per cell, $B_0 = 0.25$~T, and
$T_e = T_i$; we denote the ion cyclotron frequency
$\Omega_i = eB_0/m_i$, the cyclotron period $\tau_{ci} = 2\pi/\Omega_i$,
and the thermal ion gyroradius $\rho_i = d_i\sqrt{\beta/2}$. For
parallel propagation ($\vB_0$ aligned with the domain) the plasma has
$m_i = 100\,m_e$, $v_A/c = 10^{-4}$, and $\beta = 0.01$ on a domain of
length $102.4\,d_i$, and the circularly polarized fluctuations
$B_{R/L} = B_x \pm i B_y$ are recorded; for perpendicular propagation
($\vB_0$ across the domain) the plasma has $m_i = 400\,m_e$,
$v_A/c = 10^{-3}$, and $\beta = 0.1$ on a domain of length $192\,d_i$,
and $E_z$ is recorded. Field values are sampled along the domain at
every step, and $\omega$--$k$ spectra are obtained by a two-dimensional
Fourier transform in space and time. Both runs retain the small
resistivity of the reference setup ($\eta = 10^{-7}~\Omega$\,m)
\cite{Groenewald2023} and use the implicit advance at $\theta = 1/2$.

In the parallel case, Fig.~\ref{fig:dispersion}(a), the fluctuation
power tracks the right- and left-handed branches across the whistler
range, with the expected accumulation and thermal broadening at the
ion-cyclotron resonance. In the perpendicular case,
Fig.~\ref{fig:dispersion}(b), the spectrum reproduces the fast
magnetosonic/extraordinary branch and a ladder of ion Bernstein
harmonics up to $\omega \simeq 7\,\Omega_i$, in agreement with the
dispersion curves of Ref.~\cite{Munoz2018} and with the
explicit-solver results of Ref.~\cite{Groenewald2023}. The
Bernstein ladder is a demanding test of the ion kinetics, since the
harmonic structure arises entirely from finite-Larmor-radius effects.
The perpendicular case was run at a smaller time step than the parallel
case; the limit was set by convergence of the Newton iteration for the
higher-frequency perpendicular modes rather than by numerical stability.

The energy behavior of the two advances is compared in
Fig.~\ref{fig:expimp} for the parallel configuration with the
resistivity removed. This configuration is a stringent test of energy behavior: with
$\eta = \etah = 0$ and no external drive, every source in the energy law
of Sec.~\ref{sec:model:energy} vanishes or is reversible, so at
$\theta = 1/2$ \cref{eq:energy_law} predicts exact conservation. The
reference explicit configuration \cite{Groenewald2023} runs this problem
with a small but finite resistivity; here it is removed deliberately to
isolate the conservation properties of the two schemes. At early times
[Figs.~\ref{fig:expimp}(a) and \ref{fig:expimp}(c)] the two solvers
produce equivalent spectra. By $28\,\tau_{ci}$, however, the explicit
run is dominated by grid-scale fluctuation energy that obliterates the
physical branches, Fig.~\ref{fig:expimp}(b), while the implicit
spectrum is unchanged, Fig.~\ref{fig:expimp}(d). The energy histories, Fig.~\ref{fig:expimp}(e), show the mechanism: the explicit total energy
grows secularly and reaches nineteen times the initial energy within
$28\,\tau_{ci}$, whereas the implicit error remains below $10^{-13}$,
independent of the $10^{-6}$ solver tolerance: the cancellation of
Sec.~\ref{sec:model:energy} is algebraic in the converged iterate, so
the conservation error is set by floating-point roundoff rather than by
the tolerance of the nonlinear solve. Explicit runs at larger steps of
$1.25\times10^{-3}$, $2.5\times10^{-3}$, and $5\times10^{-3}\,\tau_{ci}$
terminate at $0.27$, $0.13$, and $0.09~\mu$s respectively. We emphasize
that the explicit field advance is sub-stepped and therefore
whistler-CFL stable; the growth is the numerical heating of the explicit
hybrid advance at zero resistivity, which in practice is suppressed by
adding artificial dissipation \cite{Groenewald2023}. The implicit
midpoint advance requires none.

Finally, Fig.~\ref{fig:theta} quantifies the numerical dissipation
introduced by $\theta > 1/2$. A single whistler mode at $k d_i = 2.0$ is initialized in the uniform
plasma of the parallel EM-mode benchmark, and
its complex amplitude is tracked as the time step is varied, mapping the
per-step damping as a function of $\omega\dt$. The measured damping follows the analytic
amplification factor of the $\theta$ scheme over the full sweep,
ramping as $(\theta - 1/2)(\omega\dt)^2$, while at $\theta = 1/2$ the
measured rate remains at the weak physical damping of the mode for all
time steps, including $\omega\dt > 1$ where the oscillation is resolved
by only a few steps per period. The $\theta$ damping is therefore a
selective dissipation of temporally under-resolved fluctuations rather
than a uniform drag, in direct analogy with the fully electromagnetic
$\theta$-PIC scheme \cite{Angus2022}, and it vanishes identically at the
energy-conserving midpoint value.

The same sweep measures the temporal accuracy of the advance and
decomposes its truncation error [Fig.~\ref{fig:theta}]: the
$\mathcal{O}(\dt)$ error term of the scheme for $\theta > 1/2$ is
purely dissipative and appears as the numerical damping rate
$\gamma_\mathrm{num} = (\theta - 1/2)\,\omega^2\dt$ of panel~(a), while
the phase error, panel~(b), is second order in $\dt$ for all $\theta$,
with the analytic coefficient $[\theta^3 + (1-\theta)^3]/3$. At
$\theta = 1/2$ the dissipative term vanishes identically and the
advance is second-order accurate.

\subsection{\label{sec:results:fgi}Cold-beam finite-grid instability}

Quasineutral hybrid PIC is subject to a finite-grid instability distinct
from that of fully kinetic PIC: a cold drifting ion population is
unstable to aliasing-driven coupling between the deposited moments and
the fields, and heats numerically until the ion temperature saturates
near the trapping level \cite{Rambo1995,StanierChacon2019}. The test
follows Fig.~1 of Stanier \emph{et al.}~\cite{StanierChacon2019}: a
uniform proton population with $T_i/T_e = 1/600$ drifts at
$u_0 = 0.1\,C_s$ along a 1D periodic domain of 40 cells, with
$n_0 = 10^{20}$~m$^{-3}$, adiabatic electrons ($\gamma = 5/3$), linear
particle shape, 200 particles per cell, $\eta = \etah = 0$, the
collocated mesh, and $\dt = 0.022\,\Delta z / C_s$, run for 4000 steps
($t = 88\,\Delta z / C_s$).

\begin{figure*}[htb]
\includegraphics[width=\textwidth]{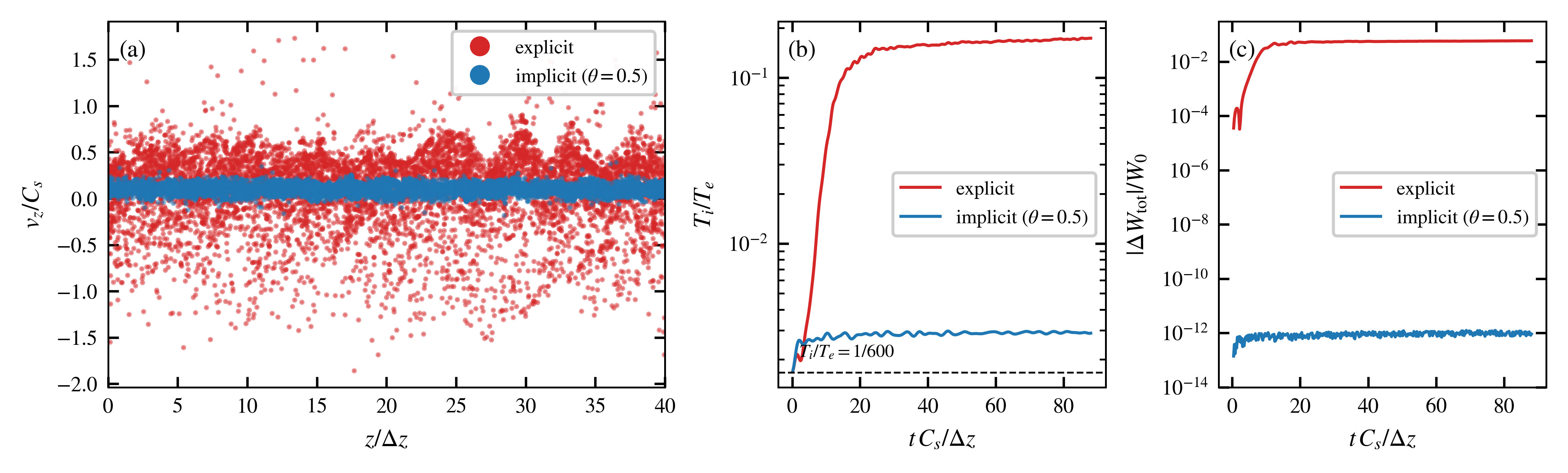}
\caption{\label{fig:fgi}Cold-beam finite-grid instability, following
Fig.~1 of Ref.~\cite{StanierChacon2019} ($T_i/T_e = 1/600$,
$u_0 = 0.1\,C_s$, linear particle shape, $\dt = 0.022\,\Delta z/C_s$,
$\eta = \etah = 0$). (a)~Ion phase space at step 4000
($t = 88\,\Delta z/C_s$) for the explicit and implicit ($\theta = 1/2$)
advances. (b)~Ion-to-electron temperature ratio; the dashed line marks
the initial value. (c)~Magnitude of the relative change in total
energy. The explicit advance develops the instability and heats the
beam to $T_i/T_e = 0.17$ with a $6\times10^{-2}$ energy gain; the
implicit advance conserves energy to machine precision and the
instability does not develop. Total momentum is conserved exactly by
both advances.}
\end{figure*}

Figure~\ref{fig:fgi} compares the two advances. With the explicit
advance the instability develops from the particle noise and heats the
beam by two orders of magnitude, to $T_i/T_e = 0.17$, with the phase
space showing the characteristic trapping vortices; the total energy
grows by $6\times10^{-2}$ of its initial value. With the implicit
advance at $\theta = 1/2$ the same configuration conserves energy to
$9\times10^{-13}$ and the instability does not develop: the beam remains
within a factor of two of its initial temperature over the full run,
carrying only the steady grid-scale ion-acoustic fluctuations sustained
by the particle noise. Total momentum is conserved exactly by both
advances. The suppression of the instability by the energy-conserving
implicit advance is consistent with the behavior reported for the
conservative implicit hybrid scheme of
Ref.~\cite{StanierChacon2019}.

\subsection{\label{sec:results:bennett}$m=0$ instability of a Bennett pinch in RZ geometry}

\begin{figure*}[htb]
\includegraphics[width=\textwidth]{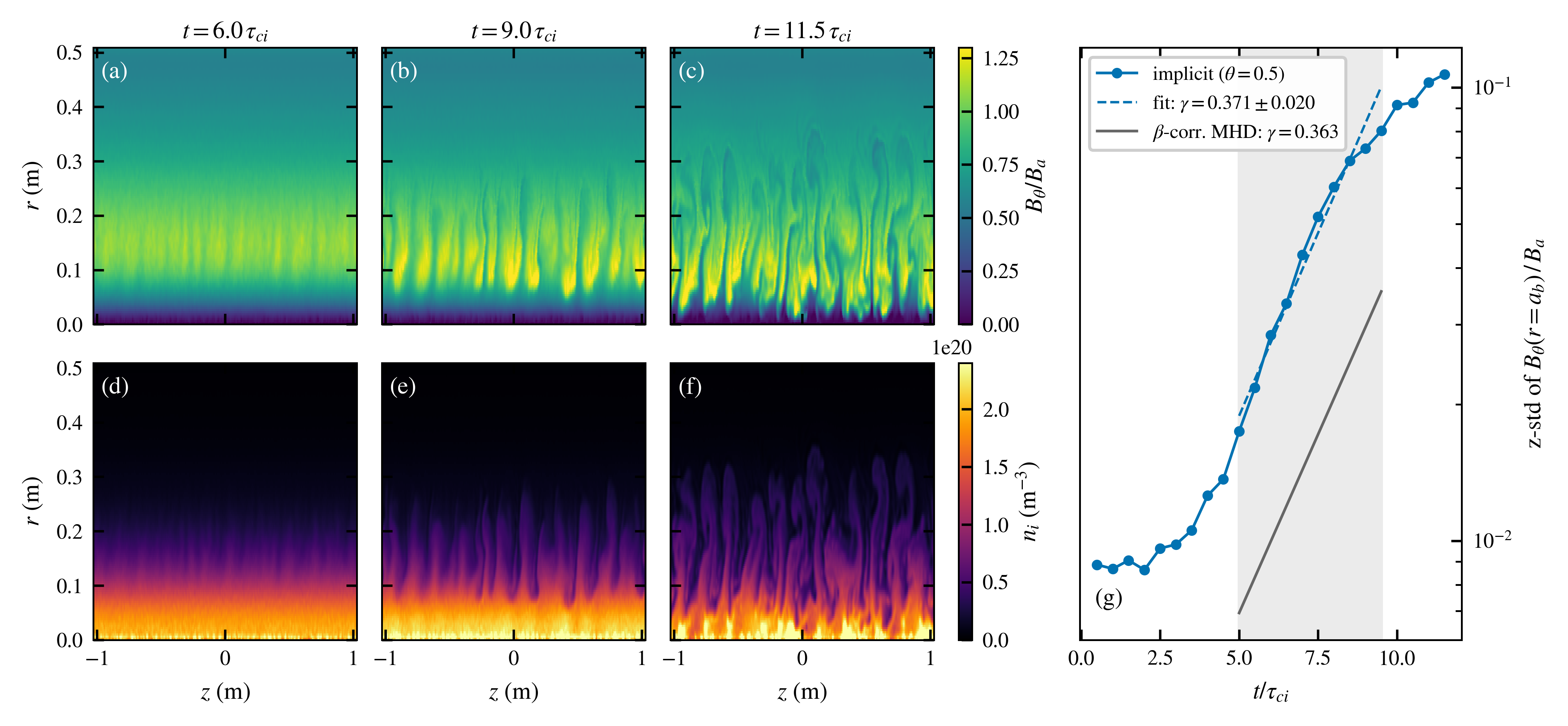}
\caption{\label{fig:bennett}$m=0$ sausage instability of a Bennett pinch
in RZ geometry, implicit solver ($\theta = 1/2$,
$\dt = 5\times10^{-3}\,\tau_{ci}$). (a--c)~$B_\theta/B_a$, with $B_a$ the
equilibrium edge field, and (d--f)~ion density at $t = 6$, $9$, and
$11.5\,\tau_{ci}$, spanning the linear phase through nonlinear
saturation. (g)~z-standard deviation of $B_\theta$ at $r = a_b$,
normalized to $B_a$. The dashed line is the exponential fit over the
shaded window, $\gamma = 0.371 \pm 0.020\,/\tau_{ci}$; the gray segment
shows the $\beta$-corrected MHD rate
$\gamma = (v_A/a_b)/\sqrt{1+\beta/2} = 0.363\,/\tau_{ci}$ as a slope
reference.}
\end{figure*}

The RZ implementation is verified on the $m=0$ (sausage) instability of a
Bennett pinch. The equilibrium is the self-pinched column
\begin{equation}
\nel(r) = \frac{n_a}{\left[1 + (r/a_b)^2\right]^{2}},
\qquad
B_\theta(r) = \frac{\mu_0 I}{2\pi}\,\frac{r}{a_b^2 + r^2},
\label{eq:bennett}
\end{equation}
with the total current $I$ set by the Bennett relation
$\mu_0 I^2 = 8\pi N_\ell k_B (T_e + T_i)$, where $N_\ell$ is the line
density. The parameters are $n_a = 2\times10^{20}$~m$^{-3}$,
$T_i = T_e = 500$~eV, and pinch radius $a_b = 10\,d_i$
($a_b = 10\,\rho_i$ at this $\beta$). Protons are loaded from
\cref{eq:bennett} with a uniform axial drift carrying the full current,
and $B_\theta$ is initialized analytically. The domain,
$32\,d_i \times 128\,d_i$ with $\Delta r = \Delta z = 0.25\,d_i$ and
periodic in $z$, is resolved with $128\times512$ cells and 250 particles
per cell; the run uses $\eta = 10^{-5}~\Omega$\,m, grid-scale
hyper-resistivity, $\theta = 1/2$, and $\dt = 5\times10^{-3}\,\tau_{ci}$,
which exceeds the whistler-stable step \cref{eq:whistler_cfl} at the
density floor by an order of magnitude.

The current-carrying column is unstable to $m=0$ perturbations, which
grow from the particle noise without external seeding.
Figure~\ref{fig:bennett}(a--f) shows the development: ripples appear in
$B_\theta$ at the pinch edge, steepen into the characteristic
density spikes and necks, and saturate with order-unity density
corrugation by $t \simeq 11\,\tau_{ci}$. The mode amplitude is tracked
by the axial standard deviation of $B_\theta$ at $r = a_b$
[Fig.~\ref{fig:bennett}(g)], which grows exponentially over roughly one
decade before rolling over into saturation.

The fitted growth rate is $\gamma = 0.371 \pm 0.020\,/\tau_{ci}$; the
fit window spans the interval between clear emergence of the mode from
the particle noise and the onset of saturation, and the fitted rate
varies by roughly $\pm 15\%$ with window placement inside the growth
phase. For comparison, the sharp-boundary MHD estimate with the
finite-$\beta$ correction, $\gamma = (v_A/a_b)/\sqrt{1 + \beta/2}$,
gives $0.363\,/\tau_{ci}$ for these parameters. Exact agreement is not
expected: the Bennett profile is diffuse rather than sharp-boundary, the
resistivity is finite, and the ion Larmor radius is only ten times
smaller than the pinch radius. An explicit reference run of the same
configuration gives $\gamma = 0.406 \pm 0.019\,/\tau_{ci}$ over the same
window, and reaches a statistically equivalent saturated state.

\subsection{\label{sec:results:frc}Three-dimensional FRC simulations: explicit vs implicit}

Field-reversed configuration (FRC) plasmas are a class of compact
high-$\beta$ toroidal magnetic systems that have been the subject of
substantial research for fusion energy
applications~\cite{Tuszewski1988,Steinhauer2011}. Despite the robustness of
FRCs observed in experiments~\cite{Slough1993}, several instability modes can
lead to the loss of magnetic confinement, including the $(m=1,n=1)$ tilt,
which manifests in the bulk plasma as either an internal or external mode
whose growth rate generally follows the MHD estimate
$\gamma_\mathrm{MHD} \approx 2 v_A / L_s$, where $L_s$ is the axial separatrix
length and $v_A$ is the characteristic Alfv\'{e}n velocity. The onset of these
modes, however, is in practice sensitive to finite-Larmor-radius effects and to
numerical errors, which can stabilize or destabilize the equilibrium,
respectively. To determine whether the proposed method reproduces the
instability behavior exhibited by a well-resolved explicit simulation, we
examine the evolution of an equilibrium FRC and track the tilt mode growth over
time, following Groenewald \textit{et al.}~\cite{Groenewald2023}. The tilt mode
amplitude is obtained from a two-dimensional Fourier transform of the
symmetric axial ion velocity $u_z^+$,
\begin{equation}
    u_z^+(z,\theta) = \int_{0}^{r_s}
      \left[u_z(r,\theta,z) + u_z(r,\theta,-z)\right] dr ,
    \label{eq:uzplus}
\end{equation}
with $u_z = J_{iz}/(n_i e)$~\cite{Groenewald2023}; $A_{11}$ denotes the
amplitude of the $(m=1,n=1)$ component of that transform, i.e.\ the
coefficient with azimuthal mode number one in $\theta$ and axial mode number
one in $z$ over the axial extent of the separatrix.

An elliptical FRC solution to the Grad--Shafranov equation is used to
initialize the problem, with kinetic parameter $S^{*} = r_s/d_i = 7.9$,
geometric elongation $\mathcal{E} = L_s/(2 r_s) = 1.66$,
$T_{i0} = 4\,\mathrm{keV}$, $T_{e0} = 400\,\mathrm{eV}$, and external field
strength $B_e = 0.35\,\mathrm{T}$, where $d_i$ is the ion inertial length of
the deuterium plasma and $r_s$ is the separatrix radius. Times in this section
are reported in Alfv\'{e}n transit times $t_A = r_s/v_A$.

A $96 \times 96 \times 360$ grid with an average of ${\sim}40$ particles per
cell was used for all simulations in this section, and $\Delta t/\tau_{ci}$ was
held at $2.5\times10^{-3}$, where $\tau_{ci}$ is the ion cyclotron period. The
step is matched between the two solvers so that the comparison isolates the
time discretization. Direct current deposition and linear particle shapes were
used, together with the mass-matrix Jacobian of
Sec.~\ref{sec:solver:massmatrices} and an embedded conducting cylindrical
boundary to emulate a vessel wall. All boundaries absorb particles, and Neumann
conditions were applied to the fields on the axial boundaries. Finally, as in
Ref.~\cite{Groenewald2023}, the electrons carry the equilibrium current; this
choice is inherited from the reference configuration for a matched comparison,
and removes the equilibrium ion flow shear that may otherwise contribute to
finite-Larmor-radius stabilization at this $S^{*}$.

\begin{figure*}[t]
\includegraphics[width=\textwidth]{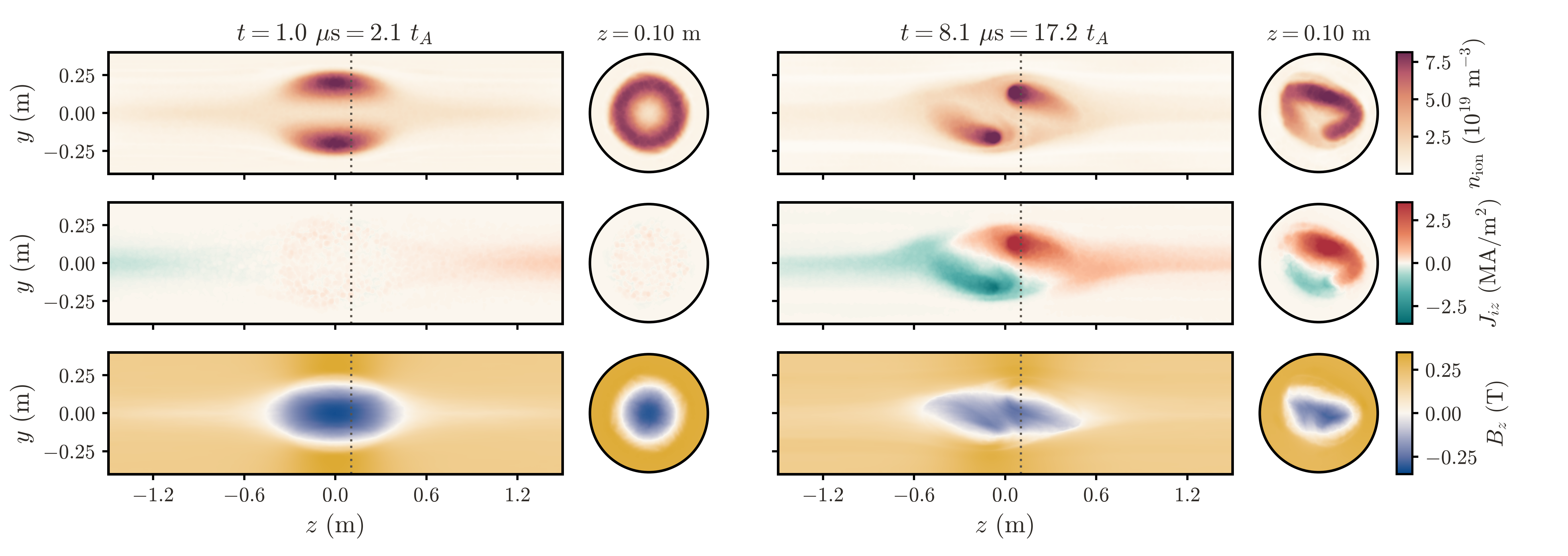}
\caption{Contours of the ion density $n_\mathrm{ion}$, axial ion current
density $J_{iz}$, and axial magnetic field $B_z$ in the $z$--$y$ and $x$--$y$
planes, at $t = 1.0\,\mu\mathrm{s}$ ($2.1\,t_A$) and
$t = 8.1\,\mu\mathrm{s}$ ($17.2\,t_A$). Implicit solver, $\theta = 0.6$.
\label{fig:frc_contours}}
\end{figure*}

\begin{figure*}[t]
\includegraphics[width=\textwidth]{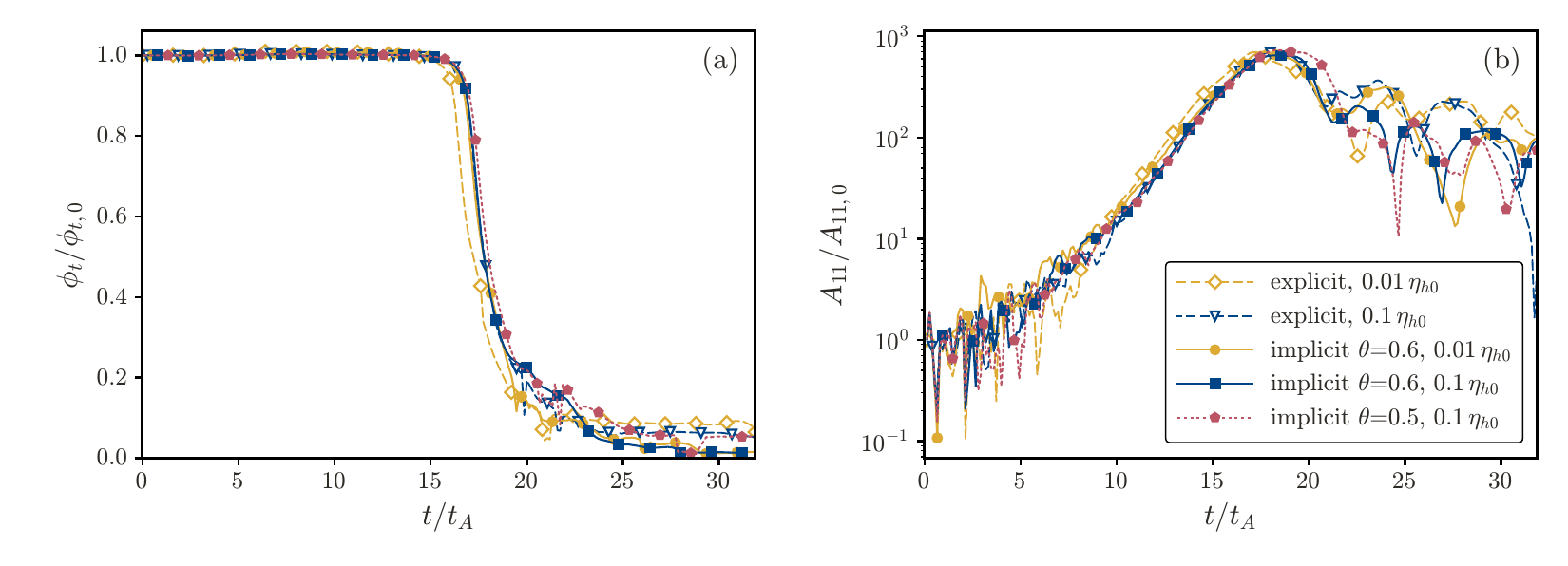}
\caption{Comparison of trapped flux $\phi_t$ (a) and tilt mode amplitude
$A_{11}$ (b), each normalized to its initial value, from the explicit and
implicit solvers at two values of the hyper-resistivity $\eta_h$ in units of
the reference value $\eta_{h0}$ of Ref.~\cite{Chacn2003}, together with an
implicit run at $\theta = 0.5$.
\label{fig:frc_comparison}}
\end{figure*}

Figure~\ref{fig:frc_contours} shows the evolution of the FRC as it transitions
from quiescence to instability using the proposed implicit solver with
$\theta = 0.6$.
The FRC is initially nominally stable, but as the tilt mode grows
exponentially the configuration changes dramatically until it is disrupted,
consistent with past results~\cite{Belova2001,Belova2003,Belova2004}.
%
High trapped flux $\phi_t$ is key to a long FRC lifetime, but instabilities
such as tilt lead to loss of confinement and hence rapid decay of the trapped
flux. Further, the stabilizing hyper-resistivity in the generalized Ohm's law
can artificially diffuse flux out of the FRC and alter the lifetime and
stability. To establish that the proposed method does not alter the instability
dynamics obtained with the explicit solver, we measure the trapped flux and the
tilt mode amplitude with both solvers at two values of $\eta_h$, spanning an
order of magnitude. Figure~\ref{fig:frc_comparison} shows the resulting traces.

The tilt mode grows out of grid noise through roughly three orders of
magnitude before saturating. 
Loss of
confinement is concurrent with that saturation rather than following it.
Least-squares fits of $\ln A_{11}$ over $t/t_A \in [12,16]$, an interval lying
between clear emergence of the mode from the noise and the onset of saturation,
give the growth rates collected in Table~\ref{tab:tilt-growth}:
$\gamma t_A = 0.50$--$0.58$, corresponding to e-folding times of
$2.0$--$1.7\,t_A$, with $R^2 \ge 0.99$ for every fit. The dominant uncertainty
is the placement of the fit window. Refitting over every sub-window of
$[12,16]$ at least $3\,t_A$ wide moves $\gamma$ by up to $\pm 0.06$; the
uncertainties quoted in Table~\ref{tab:tilt-growth} reflect this window
sensitivity.

At fixed $\eta_h$ the implicit and explicit results agree to within the window
uncertainty: at $0.1\,\eta_{h0}$ the implicit rate is $3.6\%$ below the
explicit one, small against the spread of $\pm 0.04$ obtained by moving the
window. Reducing the implicit parameter from $\theta = 0.6$ to $\theta = 0.5$
changes the rate by a further $1.8\%$, likewise within that uncertainty. The
two solvers reproduce each other's disruption and saturation times to within
about one Alfv\'{e}n transit time. For comparison, raising $\eta_h$ by an order
of magnitude changes the growth rate by $16\%$, so the sensitivity of the mode
to the physical dissipation is several times larger than its sensitivity to the
choice of solver or of $\theta$. All rates fall below the ideal-MHD estimate
$\gamma_\mathrm{MHD} = 2 v_A/L_s$, at $0.83$--$0.96\,\gamma_\mathrm{MHD}$.
After saturation the $A_{11}$ traces separate, as expected for a nonlinearly
disrupted state in which the mode amplitude is no longer a smooth function of
the initial condition.

Taken together, the agreement between the two solvers across an order of
magnitude in $\eta_h$, and the insensitivity of the growth rate to $\theta$,
indicate that the proposed method introduces no spurious disturbance that
alters the tilt mode at the level resolved by these runs. On this basis, the
method appears well suited to studies of FRC plasmas.

\begin{table}[t]
  \centering
  \caption{Linear growth rate of the $(m=1,n=1)$ tilt mode, obtained by
    least-squares fitting of $\ln A_{11}$ over $t/t_A \in [12,16]$. Rates are
    quoted per Alfv\'{e}n transit time $t_A = r_s/v_A = 0.471\,\mu\mathrm{s}$,
    with $r_s = 0.280\,\mathrm{m}$ and $v_A = 5.94 \times
    10^{5}\,\mathrm{m\,s^{-1}}$ from the initial equilibrium. Uncertainties are
    the sensitivity to fit-window placement, taken as the largest excursion of
    $\gamma$ over all sub-windows of $[12,16]$ at least $3\,t_A$ wide.
    $\eta_{h0}$ is the reference hyper-resistivity of Chac\'{o}n and
    Knoll~\cite{Chacn2003}. For comparison, the ideal-MHD tilt rate
    $\gamma_\mathrm{MHD} = 2 v_A/L_s$ gives
    $\gamma_\mathrm{MHD} t_A = 0.602$, with the separatrix length
    $L_s = 0.931\,\mathrm{m}$ measured at $t = 1.0\,\mu\mathrm{s}$, once the
    initial configuration has relaxed.}
  \label{tab:tilt-growth}\vspace*{0.2cm}
  \begin{tabular}{lccc}
    \toprule
    & \multicolumn{3}{c}{$\gamma t_A$} \\
    \cmidrule(lr){2-4}
    $\eta_h$ & explicit & implicit $\theta = 0.6$ & implicit $\theta = 0.5$ \\
    \midrule
    $0.01\,\eta_{h0}$ & $0.501 \pm 0.056$ & $0.497 \pm 0.025$ & -- \\
    $0.1\,\eta_{h0}$  & $0.580 \pm 0.019$ & $0.559 \pm 0.041$ & $0.549 \pm 0.020$ \\
    \bottomrule
  \end{tabular}
\end{table}

\section{\label{sec:performance}Solver performance}

The cost structure of the implicit advance is characterized on a single
representative configuration: a 2D force-free screw-pinch equilibrium
on $128\times128$ cells with 800 particles per cell (13M particles), on
the Yee-staggered mesh with $\theta = 1/2$, Newton and GMRES relative
tolerances of $10^{-6}$, particle suborbits enabled, and linear
particle shape with direct deposition. All timings are 100-step
averages on a single NVIDIA RTX 6000 Ada GPU in double precision.

\begin{figure}
\includegraphics[width=\columnwidth]{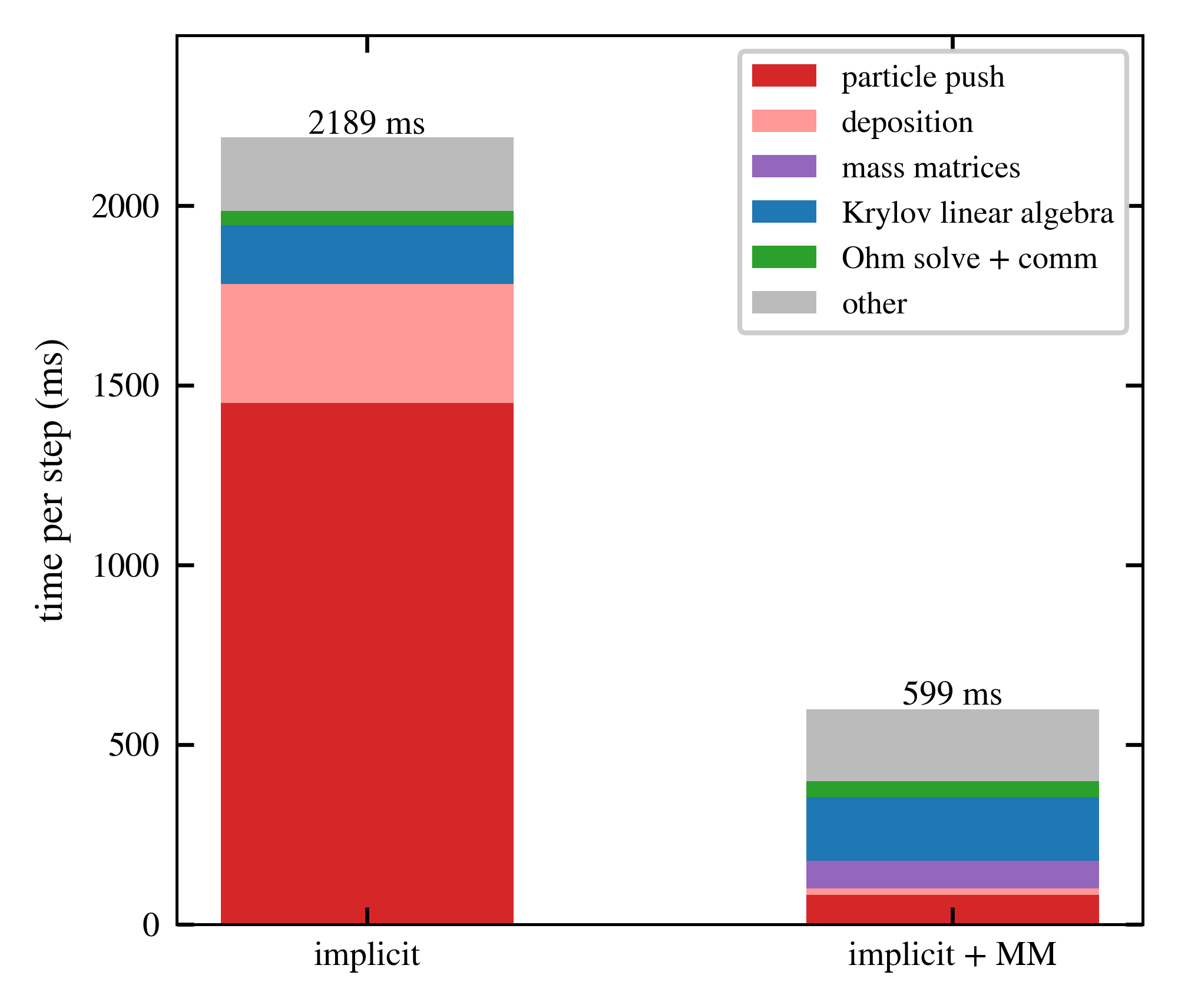}
\caption{\label{fig:performance}Wall-clock time per step of the
implicit advance, decomposed into subprocesses, without and with the
mass-matrix Jacobian ($128\times128$ cells, 800 particles per cell,
Yee mesh, $\theta = 1/2$, tolerances $10^{-6}$; 100-step averages on
one NVIDIA RTX 6000 Ada GPU). The mass matrices remove the particle
re-advance and re-deposition from the GMRES iterations, reducing the
particle work by more than an order of magnitude at the cost of a
small assembly overhead.}
\end{figure}

Figure~\ref{fig:performance} shows the wall-clock time per step
decomposed into the major subprocesses. Without mass matrices, the step
is dominated by particle operations: every GMRES iteration evaluates
the Jacobian action matrix-free, which entails a full particle
re-advance and re-deposition, and at 80 matvecs per step these account
for 81\% of the cost. With the mass-matrix Jacobian of
Sec.~\ref{sec:solver:massmatrices}, the linear-stage particle work is
replaced by a banded grid operation: the particle push and deposition
times drop by more than an order of magnitude, the mass-matrix assembly
and application add back 13\% of the (much shorter) step, and the total
cost falls by $3.7\times$. The remaining cost is dominated by the
Krylov linear algebra (chiefly the Gram--Schmidt orthogonalization),
which is purely grid-based.

The dependence of the solver cost on the time step is measured with the
single-mode configuration of Fig.~\ref{fig:theta}. Over a $40\times$
increase of $\dt$ at fixed tolerance, the Newton iteration count is
flat at 2.0 per step while the GMRES iteration count per Newton step
grows from 8.5 to 177: the nonlinear formulation remains robust far
beyond the whistler CFL limit, and the cost of large time steps is
carried entirely by the unpreconditioned Krylov solve. This identifies
preconditioning of the linear stage as the principal outstanding
optimization; it is deferred to future work. Parallel scaling is not
measured here: the residual and matvec evaluations use the same
particle and grid kernels whose multi-GPU scaling for this hybrid model
was demonstrated in Ref.~\cite{Groenewald2023}, and the Krylov
iteration adds only grid-local work and global reductions.

\section{\label{sec:summary}Summary and future work}
We have presented a $\theta$-implicit hybrid particle-in-cell algorithm,
implemented in the open-source, GPU-capable code WarpX, in which kinetic ions
are coupled to an inertialess electron fluid closed by a generalized Ohm's law
and an evolved scalar electron pressure. Fields and particles are advanced as a
single nonlinear system, reduced by nonlinear elimination of the
macro-particles and elimination of $\vB$ through the discrete Faraday law to a
fixed-point condition on $\ntheta{\vE}$ alone, \cref{eq:residual}, which is
solved by a Jacobian-free Newton--Krylov method. The time step is therefore not
constrained by the whistler CFL condition \cref{eq:whistler_cfl} and no field
sub-cycling is required; prescribed external drives enter through an exact
vector-potential splitting (Sec.~\ref{sec:model:extfields}).

Advancing the electron pressure inside the nonlinear solve, in conservative
flux form and with work terms formed from the same discrete fields at the same
mesh locations as the ion push and the Faraday update, makes the ion, magnetic,
and electron budgets cancel term by term. What remains is a single
negative-definite contribution proportional to $\phi = \theta - 1/2$,
\cref{eq:energy_law}, which vanishes identically at $\theta = 1/2$ and
otherwise damps only temporally under-resolved fluctuations. The cancellation
places no condition on the contents of Ohm's law, so any resistivity model,
hyper-resistivity, or density regularization is covered; being algebraic in the
accepted iterate, it holds to floating-point roundoff rather than to the
tolerance of the nonlinear solve.

The advance reproduces parallel and perpendicular wave dispersion including the
ion Bernstein ladder, conserves total energy to $10^{-13}$ where the explicit
advance heats secularly at zero dissipation, does not develop the cold-beam
finite-grid instability, and reproduces the $m=0$ instability of a Bennett pinch
in RZ geometry at a time step an order of magnitude beyond the whistler-stable
value. Three-dimensional simulations of a field-reversed configuration track the
$m=1,n=1$ tilt mode from grid noise through saturation. Mass matrices adapted
to the Ohm's-law Jacobian remove the particle work from the Krylov stage and
reduce the time per step by $3.7\times$; over a $40\times$ increase of $\dt$ the
Newton count remains flat at two iterations per step, with all growth in cost
carried by the unpreconditioned Krylov solve.

A number of extensions follow naturally from these results. The role of the field discretization warrants
closer study, since the collocated and staggered meshes satisfy the discrete
energy identity by different routes and differ at the grid scale, to which the
tilt onset is sensitive. A physics-based preconditioner for the Krylov stage
would remove the cost growth with $\Delta t$ identified in
Sec.~\ref{sec:performance}, and the electron closure can be extended within the
same nonlinear solve to include a conductive heat flux and bremsstrahlung
losses. The advance presented here, free of the whistler CFL stability constraint and
without secular numerical heating, together with the external-field splitting
, brings kinetic simulation of FRC formation,
acceleration, and merging over many Alfv\'en times within reach.

\section*{CRediT authorship contribution statement}
\textbf{Prabhat Kumar:} Conceptualization, Methodology, Software, Validation,
Writing -- original draft.
\textbf{William J. White:} Investigation, Formal analysis, Writing -- original draft.
\textbf{Stephen E. Clark:} Conceptualization, Methodology, Software,
Formal analysis, Writing -- review \& editing.

\section*{Declaration of competing interest}
The authors declare that they have no known competing financial
interests or personal relationships that could have appeared to
influence the work reported in this paper.

\section*{Acknowledgments}

This research used resources of the National Energy Research Scientific
Computing Center (NERSC), a Department of Energy Office of Science User Facility
using NERSC award FES-ERCAP32666. This research used the open-source particle-in-cell code WarpX. Primary WarpX
contributors are with LBNL, LLNL, CEA-LIDYL, SLAC, DESY, CERN, Helion Energy,
TAE Technologies, and Realta Fusion. We acknowledge all WarpX contributors.

\section*{Data availability}
Data will be made available on request.

\section*{Declaration of generative AI and AI-assisted technologies in the manuscript preparation process}
During the preparation of this work the author(s) used Claude Fable 5 (Anthropic) to assist in implementing the algorithm and refining the manuscript. After using this tool/service, the author(s) reviewed and edited the content as needed and take(s) full responsibility for the content of the published article.

\appendix

\section{\label{app:energy}Derivation of the discrete energy law}

This appendix derives the three budgets quoted in
Sec.~\ref{sec:model:energy} and their sum, \cref{eq:energy_law},
restores the surface and external-drive contributions, and discusses the
mesh dependence of the cancellation. Throughout, $\ntheta{\vE}$ denotes
the full Ohm field used in the Faraday update,
$\vEstar = \ntheta{\vE} - \vD$ the push field of \cref{eq:estar}, and
$\phi \equiv \theta - 1/2$.

\subsection{Magnetic field budget}

With $\bar{\vB} \equiv (\vB^{n+1}+\vB^{n})/2$, the change in the magnetic
energy \cref{eq:energy_defs} over one step is
\begin{equation}
W_B^{n+1}-W_B^{n} = \frac{1}{\mu_0}\sum_g
 \bar{\vB}_g\cdot\left(\vB_g^{n+1}-\vB_g^{n}\right)\DV .
\label{eq:app_wb1}
\end{equation}
The time-averaged and time-biased fields are related by
\begin{equation}
\bar{\vB} = \ntheta{\vB} - \phi\left(\vB^{n+1}-\vB^{n}\right),
\label{eq:app_phi_identity}
\end{equation}
and the discrete Faraday advance gives
$\vB^{n+1}-\vB^{n} = -\dt\,\curl\ntheta{\vE}$. Substituting both into
\cref{eq:app_wb1} yields
\begin{align}
\frac{W_B^{n+1}-W_B^{n}}{\dt} ={}&
 -\frac{1}{\mu_0}\sum_g \ntheta{\vB}_g\cdot
   \left(\curl\ntheta{\vE}\right)_g\DV \nonumber\\
& - \frac{\phi}{\mu_0\,\dt}
   \sum_g \left| \vB_g^{n+1}-\vB_g^{n} \right|^2 \DV .
\label{eq:app_wb2}
\end{align}
For periodic boundaries the discrete curl operators of both grid types are
mutually adjoint,
$\sum_g \vB\cdot(\curl\vE)\,\DV = \sum_g \vE\cdot(\curl\vB)\,\DV$: on the
Yee mesh this is the standard mimetic property of the staggered curl pair,
and on the collocated mesh it follows from the antisymmetry of the
centered difference. Using this identity and Amp\`ere's law
$\mu_0\ntheta{\vJ} = \curl\ntheta{\vB}$ in \cref{eq:app_wb2} gives
\cref{eq:wb_budget}.

\subsection{Ion budget}

The velocity update in \cref{eq:particle_cn} dotted with
$m_p\bar{\vv}_p$ gives, per particle,
\begin{equation}
\frac{m_p}{2}\left(|\vv_p^{n+1}|^2 - |\vv_p^{n}|^2\right)
 = q_p\,\dt\;\bar{\vv}_p\cdot\vEstar(\bar{\vx}_p),
\label{eq:app_wi1}
\end{equation}
since $\bar{\vv}_p\cdot(\bar{\vv}_p\times\ntheta{\vB}_p) = 0$ exactly.
Writing the gathered field as
$\vEstar(\bar{\vx}_p) = \sum_g S^E_{gp}(\bar{\vx}_p)\,\vEstar_{g}$ and
the deposited current as
${\nhalf{\vJi}}_{g} = \DV^{-1}\sum_p q_p \bar{\vv}_p\, S^J_{gp}(\bar{\vx}_p)$,
the sum of \cref{eq:app_wi1} over particles collapses onto the grid when
$S^E_{gp} = S^J_{gp}$, giving \cref{eq:wi_budget}. The requirement is that
the gather and deposit use the same shape function per field component at
the same (time-centered) particle positions; on staggered grids this is
realized with the momentum-conserving field interpolation.

\subsection{Electron budget}

Summing \cref{eq:pe_advance} over the grid, the conservative flux form of
the transport term telescopes to zero on a periodic mesh, leaving
\begin{equation}
\sum_g \frac{\ntheta{\pe} - \pe^{n}}{\theta\,\dt}\,\frac{\DV}{\gamma-1}
 = \sum_g \left( \ntheta{\vE}\cdot\vJe
 + \vD\cdot\nhalf{\vJi} \right)_g \DV .
\label{eq:app_we1}
\end{equation}
The extrapolation \cref{eq:extrap} gives the exact identity
$(\pe^{n+1}-\pe^{n})/\dt = (\ntheta{\pe}-\pe^{n})/(\theta\,\dt)$, so the
left-hand side equals $(W_e^{n+1}-W_e^{n})/\dt$ and \cref{eq:we_budget}
follows.

\subsection{Assembly}

Adding \cref{eq:wi_budget,eq:wb_budget,eq:we_budget} and abbreviating the
grid sum with volume weight as $\langle\cdot\rangle$,
\begin{align}
&\frac{(W_i{+}W_B{+}W_e)^{n+1}-(W_i{+}W_B{+}W_e)^{n}}{\dt} \nonumber\\
&\quad = \left\langle \nhalf{\vJi}\cdot\vEstar \right\rangle
 - \left\langle \ntheta{\vJ}\cdot\ntheta{\vE} \right\rangle
 + \left\langle \ntheta{\vE}\cdot\vJe + \vD\cdot\nhalf{\vJi} \right\rangle
 \nonumber\\
&\qquad - \frac{\phi}{\mu_0\,\dt}
   \sum_g \left| \vB_g^{n+1}-\vB_g^{n} \right|^2 \DV .
\label{eq:app_sum}
\end{align}
Substituting $\vEstar = \ntheta{\vE} - \vD$ in the first term and
$\vJe = \ntheta{\vJ} - \nhalf{\vJi}$ in the third,
\begin{equation}
\label{eq:app_cancel}
\begin{split}
&\left(\langle \bm{J}_i^{n+1/2}\!\cdot\bm{E}^{n+\theta}\rangle
      -\langle \bm{E}^{n+\theta}\!\cdot\bm{J}_i^{n+1/2}\rangle\right)\\
&\quad{}+\left(\langle \bm{D}\cdot\bm{J}_i^{n+1/2}\rangle
      -\langle \bm{J}_i^{n+1/2}\!\cdot\bm{D}\rangle\right)\\
&\quad{}+\left(\langle \bm{E}^{n+\theta}\!\cdot\bm{J}^{n+\theta}\rangle
      -\langle \bm{J}^{n+\theta}\!\cdot\bm{E}^{n+\theta}\rangle\right)=0,
\end{split}
\end{equation}
and \cref{eq:energy_law} follows. Note that the cancellation is algebraic
in the grid quantities: it requires the products in
\cref{eq:wi_budget,eq:wb_budget,eq:we_budget} to be formed from the same
discrete fields at the same locations, but places no condition on the
functional form of Ohm's law, the resistivity model, or any regularization
inside $\ntheta{\vE}$ and $\vD$.

\subsection{Boundary and external-drive contributions}

For non-periodic boundaries, two surface terms appear. The discrete
integration by parts behind \cref{eq:wb_budget} produces the Poynting flux
of $\ntheta{\vE}\times\ntheta{\vB}/\mu_0$ through the domain surface, and
the flux form of the pressure transport in \cref{eq:pe_advance} produces
the electron enthalpy flux $\gamma\,\pe\vVe/(\gamma-1)$ through the same
surface. With the external drive of Sec.~\ref{sec:model:extfields}, the
particles are pushed with $\vEstar + \vE_\mathrm{ext}$, adding the input
power $\langle \nhalf{\vJi}\cdot\vE_\mathrm{ext}\rangle$ to the ion budget;
the energy of the prescribed pair
$(\vE_\mathrm{ext},\vB_\mathrm{ext})$ is bookkept externally. These terms
are physical fluxes and inputs, not numerical residuals, and
\cref{eq:energy_law} holds for the closed part of the system with them
subtracted.

\subsection{Mesh dependence of the cancellation}

On the collocated grid all fields share the nodal mesh, so
$\ntheta{\vE}$, $\vD$, $\nhalf{\vJi}$, and $\ntheta{\vJ}$ in
\cref{eq:app_sum} denote the same nodal values in every budget and the
cancellation \cref{eq:app_cancel} is pointwise. On the Yee-staggered mesh
each vector component lives at its own location, and every product in the
three budgets is formed per component at that location: the
gather/deposit identity behind \cref{eq:wi_budget} holds per component,
the adjointness sum behind \cref{eq:wb_budget} places
$\ntheta{\vJ}\cdot\ntheta{\vE}$ at the same component locations, and the
electron bracket of \cref{eq:pe_advance} is evaluated there and split to
the pressure nodes with volume weights chosen so that the grid sum
reproduces the component-located sum exactly. The cancellation
\cref{eq:app_cancel} is then exact per component, and in Cartesian
geometry \cref{eq:energy_law} holds on the staggered mesh as on the
collocated mesh.

In cylindrical geometry the radial metric enters both the discrete
adjointness identity and the component-to-node splitting through the cell
volumes (including the axis cell, whose nodal control volume is
$\pi\,\Delta r^2 \Delta z/4$ and for which the radial flux $rF_r$
vanishes at $r = 0$), and the two weightings agree at interpolation order
rather than identically. The resulting residual in \cref{eq:energy_law}
is independent of the solver tolerance; it is measured in
Sec.~\ref{sec:results} and is small compared to the physical energy
exchanges in all cases tested.

We note that the decomposition of the electron work into Hall,
frictional, and pressure channels, used in Sec.~\ref{sec:model:energy}
for physical interpretation, is nowhere required by the derivation: the
orthogonality $\vJe\cdot(\vJe\times\vB) = 0$ holds pointwise only where
all three vector components share a location, so on staggered meshes the
attribution to individual channels (though not their sum) is defined at
the interpolation order of the staggered fields.

\bibliographystyle{elsarticle-num}
\bibliography{references}

\end{document}